\documentclass[prl,twocolumn]{revtex4}

\usepackage{amsmath}
\usepackage{amssymb}
\usepackage{amsfonts}
\usepackage{color}
\usepackage{graphicx}

\newcommand{\RR}{\right}
\newcommand{\LL}{\left}
\newcommand{\m}{\mathrm}

\newcommand{\PauliX}{\sigma_x}
\newcommand{\PauliY}{\sigma_y}
\newcommand{\PauliZ}{\sigma_z}

\begin{document}

\title{Hybrid circuit cavity quantum electrodynamics with a micromechanical resonator}


\author{J.-M.~Pirkkalainen}
\author{S.~U.~Cho}
\thanks{Present address: Korea Research Institute of Standards and Science, Daejeon 305-340, Republic of Korea}\author{Jian~Li}
\author{G.~S.~Paraoanu}
\author{P.~J.~Hakonen}
\author{M.~A.~Sillanp\"a\"a}
\thanks{Present address: Department of Applied Physics, Aalto University School of Science, P.O. Box 11100, FI-00076 Aalto, Finland}

\affiliation{O.~V.~Lounasmaa Laboratory, Low Temperature Laboratory, Aalto University, P.O. Box 15100, FI-00076 Aalto, Finland.}

\begin{abstract}
Hybrid quantum systems with inherently distinct degrees of freedom play a key role in many physical phenomena. Famous examples include cavity quantum electrodynamics \cite{HarocheRMP}, trapped ions \cite{WinelandRMP}, or electrons and phonons in the solid state. Here, a strong coupling makes the constituents loose their individual character and form dressed states. Apart from fundamental significance, hybrid systems can be exploited for practical purpose, noteworthily in the emerging field of quantum information control. A promising direction is provided by the combination between long-lived atomic states \cite{WinelandRMP,DeMille} and the accessible electrical degrees of freedom in superconducting cavities and qubits \cite{Nakamura,ClarkeReview}. Here we integrate circuit cavity quantum electrodynamics \cite{Wallraff_Nature_2004,Delftqed} with phonons. Besides coupling to a microwave cavity, our superconducting transmon qubit \cite{transmon} interacts with a phonon mode in a micromechanical resonator, thus representing an atom coupled to two different cavities. We measure the phonon Stark shift, as well as the splitting of the qubit spectral line into motional sidebands, which feature transitions between the dressed electromechanical states. In the time domain, we observe coherent conversion of qubit excitation to phonons as sideband Rabi oscillations. This is a model system having potential for a quantum interface, which may allow for storage of quantum information in long-lived phonon states, coupling to optical photons, or for investigations of strongly coupled quantum systems near the classical limit.
\end{abstract}

\maketitle


Superconducting quantum bits based on Josephson junctions \cite{ClarkeReview} have offered an unparalleled testing ground for quantum mechanics in relatively large systems. At the same time, Josephson devices constitute a promising implementation for quantum information processing. Basic quantum algorithms have indeed been recently demonstrated with phase \cite{Mariantoni07102011} and transmon \cite{SchoelkopfError,WallraffToff,EsteveSpeed} qubits. The latter operate in the circuit cavity quantum electrodynamics (QED) architechture, in which the qubits couple to an on-chip \cite{Wallraff_Nature_2004} or 3-dimensional microwave cavity resonator \cite{Paik_PRL_2011}. The circuit QED setup, which enables coupling of qubits and non-destructive measurements of quantum states, can be regarded as the most feasible platform for quantum information.

The forthcoming challenges in circuit	QED include the construction of an interface to the storage and retrieval of qubit states in a long-lived quantum memory, as well as quantum communication \cite{Zeilinger} between spatially separated superconducting qubits. Hybrid quantum systems are showing promise for these goals because in principle one can combine the specific assets of each ingredient. Merger of macroscopic qubits with spin ensembles is intriguing due to the long lifetime of the latter \cite{Zhu_Nature_2011,Kubo_PRL_2011}, but with the drawback of a difficult access and small coupling at the level of a single atomic degree of freedom.

\begin{figure}
 \includegraphics[width=0.75\linewidth]{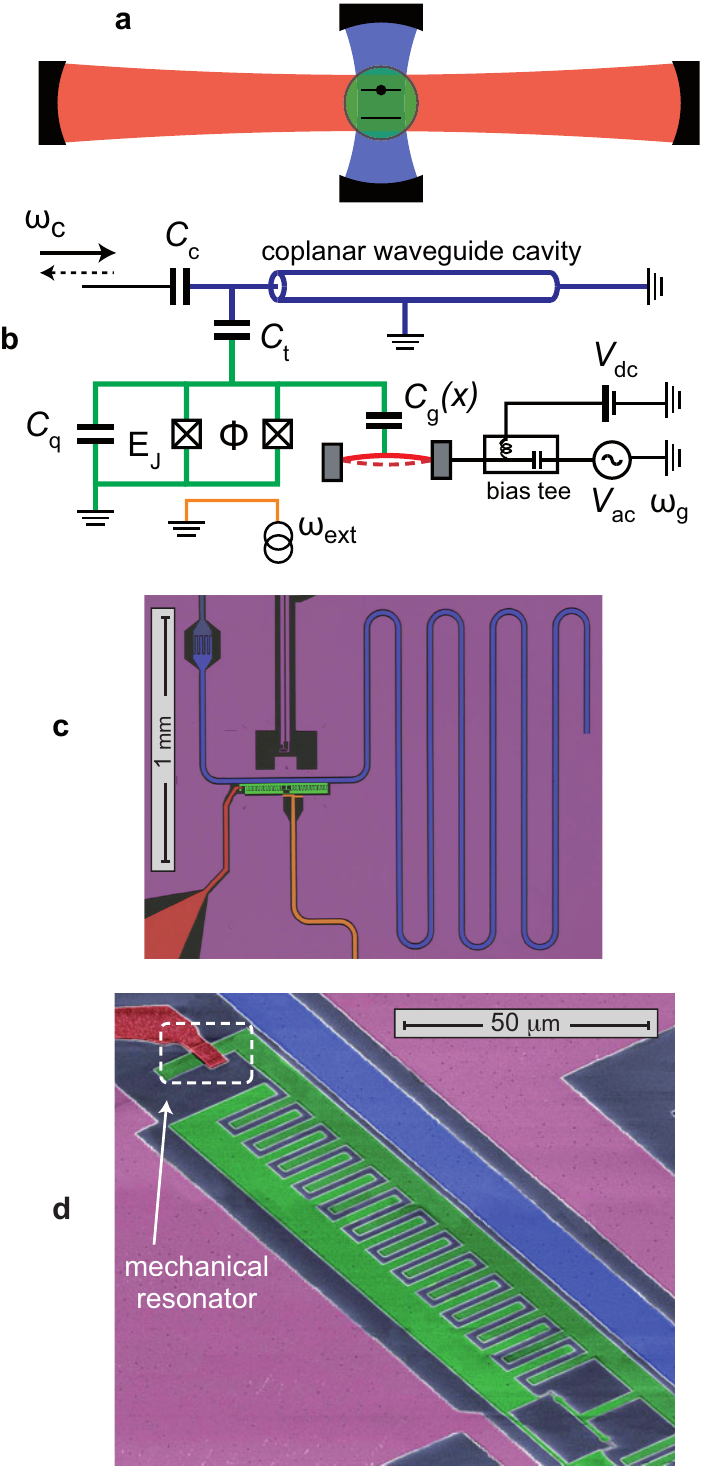}
  \caption{\textbf{Hybrid QED setup}. \textbf{a}, Illustration of a cavity quantum electrodynamics (QED) system consisting of an atom (green) coupled to two cavities. Blue color corresponds to the (photonic) microwave resonator and red to the (phononic) lower-frequency mechanical resonator. \textbf{b}, Circuit schematics of an analogous electromechanical tripartite system, measured via a microwave cavity with a probe tone at frequency $\omega_c$. \textbf{c}, Optical image of the chip. Shown is the quarter-wavelength microwave cavity, the superconducting transmon qubit playing the role of the atom, the mechanical resonator with its external control, a flux control line (orange), and the ground plane (magenta). \textbf{d}, Scanning electron micrograph showing the $5 \, \mu$m long and $4 \, \mu$m wide bridge-type mechanical resonator (dashed box) suspended above the qubit island.}
  \label{fig:scheme}
\end{figure}

Micromechanical resonators were brought to the quantum regime of their motion only very recently \cite{Teufel_Nature_2011,Chan_Nature_2011}. They have been suggested as a plausible interfacing medium for Josephson junction qubits \cite{Armour2002,TianPRB2005,Etaki2008}, with a potential to reach the aforementioned targets. The first experiment of this kind demonstrated the interaction between a charge qubit and a beam resonating at the frequency $\omega_m/2\pi \sim 60$ MHz \cite{LaHaye_Nature_2009}. This experiment, using a read-out via the micromechanical device, did not allow for time-resolved measurements, a pre-requisite for coherent state transfer. The resonant coupling was recently demonstrated with a microwave-regime piezoelectric device \cite{OConnel_Nature_2010}, having $\omega_m/2\pi \sim 6.2$ GHz, but the energy decay times were short due to high mechanical frequency and special materials.

Our approach to improve on these issues is to utilize a membrane-type micromechanical resonator embedded in a complete circuit QED device. This allows for access to the complete toolbox of circuit QED. Notwithstanding our mechanical resonator can be understood classically, our aim is also to further the understanding of the emergence of ultrastrong coupling of a micromechanical phonon degree of freedom, representing matter, to microwave light.


On the conceptual level, our device is analogous to an optical cavity QED system where a two-level atom is coupled to two cavities having different frequencies (see Fig.~1a). We use a superconducting transmon qubit \cite{transmon}, with the capacitance $C_q$, single-electron charging energy $E_C \simeq e^2/2C_q$ and Josephson energy $E_J$ as an artificial two-level system. It is coupled to a highly detuned phononic cavity formed by a suspended aluminum membrane (Fig.~1d) with $\omega_m/(2\pi) \sim 72$ MHz by means of a position-dependent capacitance $C_g(x)$, as well as to a nearly resonant photonic cavity realized as an on-chip coplanar waveguide microwave resonator with frequency $\omega_c$.

The equivalent electrical circuit (Fig.~1b) for the two-resonator circuit QED system allows writing the Hamiltonian
\begin{align}
    \Hat{H} = & \hat{H}_{\mathrm{q}} + \hbar \omega_c (\hat{a}^{\dagger} \hat{a} + 1/2) + \hbar \omega_m (\hat{b}^{\dagger} \hat{b} + 1/2) \nonumber \\
    & + \hbar g_c (n_{0} - \hat{n}) ( \hat{a}^{\dagger} + \hat{a}) + \hbar g_m ( n_{0} - \hat{n}) ( \hat{b}^{\dagger} + \hat{b}), \label{eq:Hamiltonian}
\end{align}
where $\hat{H}_{\mathrm{q}}$ is the qubit Hamiltonian (see Supplementary, Eq.~(S17)), $\hat{a}$ and $\hat{b}$ are the annihilation operators of the microwave cavity and of the mechanical resonator, respectively, and $\hat{n}$ is the number operator of Cooper pairs on the transmon island. The qubit-mechanical resonator interaction is given by the electromechanical coupling energy $\hbar g_m = x_{\m{zp}}  V_{\m{dc}} \frac{d C_g }{d x} \frac{2e}{C_{q}} $. Here, $x_{\m{zp}} \sim 4$ fm is defined as the mechanical zero-point motion, $V_{\m{dc}}$ is a constant voltage applied to the mechanical resonator, and $n_{0} = C_g V_{\m{dc}} /2e$ is the dimensionless gate charge. The coupling between the qubit and the cavity, characterized by the energy $\hbar g_c$, allows for a dispersive measurement of the state of the qubit using the state-dependent pull on the cavity \cite{Blais_PRA_2004}.

\begin{figure*}
 \includegraphics[width=0.99\linewidth]{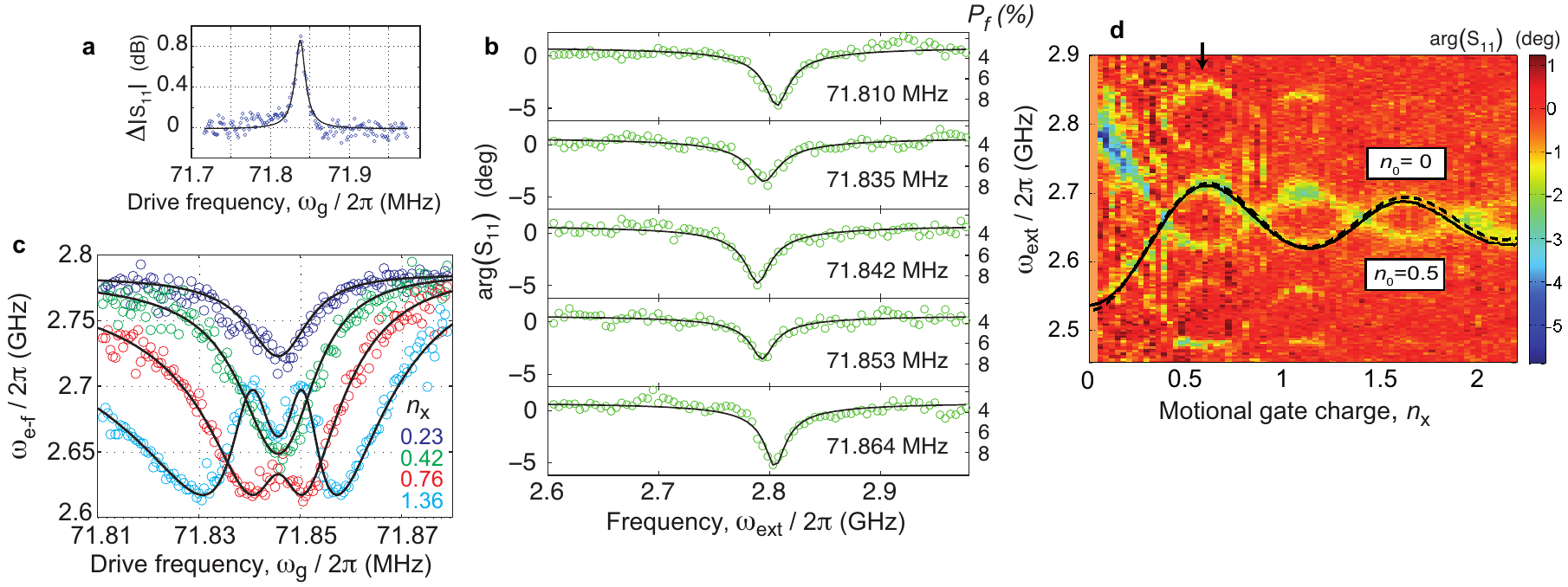}
  \caption{\textbf{Mechanical Stark shift}. \textbf{a}, The driven mechanical resonance appears as enhanced absorption in a single-tone cavity spectroscopy. \textbf{b}, In two-tone spectroscopy (flux bias $\Phi_{\m{dc}} / \Phi_0 = 0.368$, and $n_0 = 0.5$) the dip representing the transition $|e \rangle$-$|f \rangle$ (between the first and second excited states) of the transmon qubit, is observed to red-shift when the mechanical drive frequency is swept across resonance. Circles denote the measured phase of the cavity reflection, and the solid lines are Lorentzian fits. The scale on the right is the population (probability) on $|f \rangle$. With resonant mechanical drive $\omega_g/2 \pi = \omega_m/2 \pi = 71.842$ MHz, a mechanical motion is excited with an rms amplitude of 140 fm, about 30 times $x_{\m{zp}}$, and the phonon number $N_m \sim 10^3$. In gate charge units, this equals a motional gate charge $ n_x = 0.09$. \textbf{c}, Shift of the qubit transition at flux bias $\Phi_{\m{dc}} / \Phi_0 = 0.370$ as a function of driving frequency, with four drive amplitudes as labeled. Circles denote the center of the dip extracted from the data as in \textbf{b}, and the solid lines are the modeled response of a mechanical resonator with $Q_m = 5500$. \textbf{d}, Highly non-linear regime of the mechanical Stark shift with resonant drive at $n_{0} = 0$. Due to a single electron fluctuating on and off the island, also the transition line corresponding to $n_{0} = 0.5$ is visible as a mirror image. The solid line is a result of full numerical Floquet modeling (supplementary, II.G), and the dashed line is Eq.~(\ref{eq:bessel}).}
  \label{fig:stark}
\end{figure*}

We use the three lowest energy eigenstates of the transmon qubit, labeled as the ground state $ |g \rangle$, and the first and second excited states $|e \rangle$ and $|f \rangle$, respectively. The corresponding transition frequencies, tunable in the GHz regime by a flux bias $\Phi_{\m{dc}}$, are  $ \omega_{\m{g-e}}$ and  $ \omega_{\m{e-f}}$. Somewhat different from the standard concept of a qubit, which usually involves the ground state, we treat $|e \rangle$ and $|f \rangle$ as the two-level system. This is beneficial, because the phonon coupling grows for higher levels of the qubit. Also, for the same reason, we use the transmon qubit in the regime of relatively small $E_J/E_C$ ratio.

The coupling of the qubit to the phonons in the mechanical resonator in Eq.~(1) is described by a generic light-matter interaction. The quantum eigenstates of this coupled system are dressed states, which are combinations of the qubit states and the mechanical resonator Fock states $|N_m \rangle$ with phonon number $N_m$. For large $N_m$ the effective coupling $g_m\sqrt{N_m+1}$ becomes comparable to the qubit frequency, and the eigenstates attain a highly dressed character (Supplementary Fig.~S8).

The qubit transition frequencies are also expected to depend on the number of quanta (ac Stark shift). As a phonon-induced effect, it was measured indirectly through the mechanical resonator \cite{LaHaye_Nature_2009}. One can calculate the phonon (or, mechanical) Stark shift of the frequency $ \omega_{\m{e-f}}$ of the $|e \rangle$-$|f \rangle$ transition going beyond the linear regime:
\begin{equation}
\label{eq:bessel}
\begin{split}
\Delta \omega_{\m{e-f}}/2 \pi = - \frac{\epsilon_{\m{e-f}}}{2}  \cos(2 \pi n_0) \LL[J_0 (2 \pi  n_x) - 1 \RR] \, ,
\end{split}
\end{equation}
where the argument of the Bessel function is the motional gate charge amplitude $n_x = \LL( \frac{ \hbar g_m }{4 E_C} \RR) \sqrt{N_m}$, 
  and the charge dispersion of the transmon qubit is $\epsilon_{\m{e-f}} \simeq E_C \frac{2^{4m+5}}{m!}\sqrt{\frac{2}{\pi}} \LL(\frac{E_J}{ 2E_C}\RR)^{\frac{m}{2}+\frac{3}{4}} \exp\LL(-\sqrt{8E_J/E_C}\RR)$, with $m=3$ for this transition. The shift grows first linearly with the phonon number, according to $\Delta \omega_{\m{e-f}}  \simeq \frac{\epsilon_{\m{e-f}}}{2} \pi^2 \cos(2 \pi n_0) \LL( \frac{ \hbar g_m }{4 E_C}\RR)^2 N_m.$

The device (see Fig.~1c,d) was fabricated out of aluminum on a sapphire substrate in a process comprising three layers of electron-beam lithography. In order to suspend the mechanical resonator, we used organic resist as a sacrificial layer \cite{Pashkin08} (see Supplementary I.A.). The quarter-wavelength microwave cavity has a resonant frequency of $\omega_c/2 \pi = 4.84$ GHz, coupling to the qubit capacitively at the rate $g_c/2 \pi = 100$ MHz. The cavity is coupled to the measurement circuitry by a large input/output capacitor resulting in the external linewidth $\gamma_E/2\pi \simeq 15$ MHz. The device is mounted in a dilution refrigerator at a temperature of $T= 25$ mK. Although we did not access the thermal phonon number, we expect the mode to thermalize, as observed for membranes \cite{Teufel_Nature_2011}, down to the thermal occupancy $N_m^T =  k_B T /\hbar \omega_m\lesssim 10 $.


We first identified, differently from the forthcoming measurements, the lowest mechanical eigenmode of the membrane. The electromechanical interaction was switched on by applying a dc voltage $V_{\m{dc}} = 5$ V, which allows an electromechanical coupling $g_m/2\pi = 4.5$ MHz. The mechanical resonator was driven strongly by an additional ac tone combined into a bias-T inside the cryostat as shown in Fig.~1b. When the qubit is biased near the microwave cavity resonance, the cavity absorption in a single-tone measurement is altered when the mechanical mode becomes highly excited. This is due to the qubit being excited to higher levels, which causes changes in cavity pull and absorption. This way, we obtain the mechanical peak at $\omega_m/2 \pi = 71.842$ MHz, as displayed in Figure~2a. In what follows, we focus on smaller motion amplitudes.



By virtue of different frequency ranges of the qubit and the mechanical resonator, we cannot bring them on resonance by tuning the qubit frequency with a flux bias. A highly detuned regime, analogously to trapped ions \cite{WinelandRMP,Wineland1996,Blatt1999}, nonetheless allows for complete quantum control. The pertinent phenomena are the Stark shift of the qubit transition frequency, and sideband transitions.

A two-tone spectroscopy is used to investigate the qubit $|e \rangle$-$|f \rangle$  transition for detecting the Stark shift. A standard method to probe this transition is to start by applying resonant microwaves in order to excite the qubit from the ground state $|g \rangle$ to $|e \rangle$. However, it turns out that in our system a small population $P_e \sim 25$ \% exists on $|e \rangle$ even in equilibrium, thus an initial excitation is not necessary. We also similarly find $P_f \sim 5$ \%. The actual excitation microwave tone (frequency $\omega_{\m{ext}}$) is applied to the transmon flux coil, and the increased population of the level $|f \rangle$ is distinguished via the phase $\arg(S_{11})$ of the probe tone. This phase carries information on the populations $P_i$ on level $i$; here, we are interested in enhanced $P_f$ characterizing the transition frequency. A shift of $\Delta \omega_{\m{e-f}}/2 \pi =-18$ MHz in Fig.~2b from the bare value $\omega_{\m{e-f}}/2\pi = 2.81$ GHz can be attributed to the mechanical Stark shift.



We can extract the center of the $|e \rangle$-$|f \rangle$ transition dip and vary the motion amplitude, as shown in Fig.~2c. A slight difference of the bare value to that in Fig.~2b is due to a flux offset of $\sim 2$ m$\Phi_0$. Towards increasing vibration amplitude, the shift becomes more pronounced, and thereafter assumes a highly nonlinear and oscillatory behavior when the motional gate charge reaches values of the order $ n_x \gtrsim 0.5$. This rather extreme regime can be explored by fixing a resonant mechanical drive, as implemented in the measurement shown in Fig.~2d. This data shows the Bessel-type oscillatory Stark shift, in accord with Eq.~(\ref{eq:bessel}).


To model the shift, we first note that the total time-dependent gate charge is comprised of two components: $\Delta n_g = n_x + C_g V_{\m{ac}}/2e$,
where the second term is a cross talk from the ac voltage drive on the mechanical resonator. On resonance, the latter contributes 8 \% of $n_x$. Since the two terms have opposite phases when the drive is below/above $\omega_m$, they sum up to make the Stark shift asymmetric, as clearly seen in Fig.~2c. We used the motional gate charge $n_x$ at the frequency $\omega_m$ as a common fitting parameter. The analytic result in Eq.~(\ref{eq:bessel}) already gives an excellent agreement with the measurement, as shown in Fig.~2d. Beyond analytics, we made a complete numerical calculation using a Floquet modeling of the full transmon Hamiltonian (see Supplementary II.G), matching well the observed spectroscopy in Fig.~2c,d.

The excellent match to the theory clearly vindicates the mechanical origin of the Stark shift. On top of that, the highly non-linear shift can be seen as a manifestation of ultrastrong coupling from light to matter \cite{Deppe2010,BSpaper}. Indeed, the frequency shift due to coupling goes far beyond the linear regime, exceeding the bare mechanical frequency. These effects, however, can be understood by treating the mechanical resonator classically, due to high phonon number.



We can use sideband transitions to transfer quanta between the qubit and the mechanical resonator \cite{Wineland1996,Blatt1999}. More specifically, the transitions occur between the coupled eigenstates which are combinations of those of the qubit and of the mechanical resonator. The degree of this dressing depends on the parameter choice. In Fig.~3c, the effective coupling is small enough such that the description in terms of separable qubit and phonon states is valid for the present purpose. Hence, an intuitive appealing picture is to describe them by using uncoupled states $|e, N_m \rangle$, $|f, N_m' \rangle$, see Fig.~3a.
 
Since the data of Fig.~3b were taken at the gate charge extreme offsets ($n_0 = 0, \, 0.5$), we obtain only two-phonon processes due to symmetry reasons. However, offset drift during data taking rendered the first order sidebands visible in Fig.~3b. As the Rabi frequency is relatively low in Fig.~3b, we can use an analytical result for the qubit population, obtaining an excellent agreement with the measured sidebands. In Fig.~3c instead, this model which supposes that the peaks do not overlap, cannot be used. When the motional gate charge is $n_x \simeq 0.4$, the qubit frequency is insensitive to $n_0$, and hence under this situation we can eliminate the dephasing due to gate charge fluctuations, as well as the fluctuating single electron (see supplementary Figs.~S9b, S11). 

Although the red/blue sidebands are associated with cooling/heating of the mechanical resonator, at high phonon occupancy as here these effects become insignificant. Nevertheless, operations on the motional sidebands are a plausible way towards the single-phonon regime. Unless the qubit is driven on the sideband resonance, it is effectively decoupled from the mechanical resonator. This idea, complemented by the use of fast flux shifts which tune the qubit frequency, will allow for the coupling to be switched on and off over ns-timescale. We mention that the presently used charge regime allows for a relatively strong coupling, but possibly makes the qubit susceptible to background charge noise. A very plausible approach is large $E_J/E_C$ where charge noise vanishes; this requires $g_m$ a few times larger (supplementary, II.I) to facilitate the quantum regime.

\begin{figure}
 \includegraphics[width=0.9\linewidth]{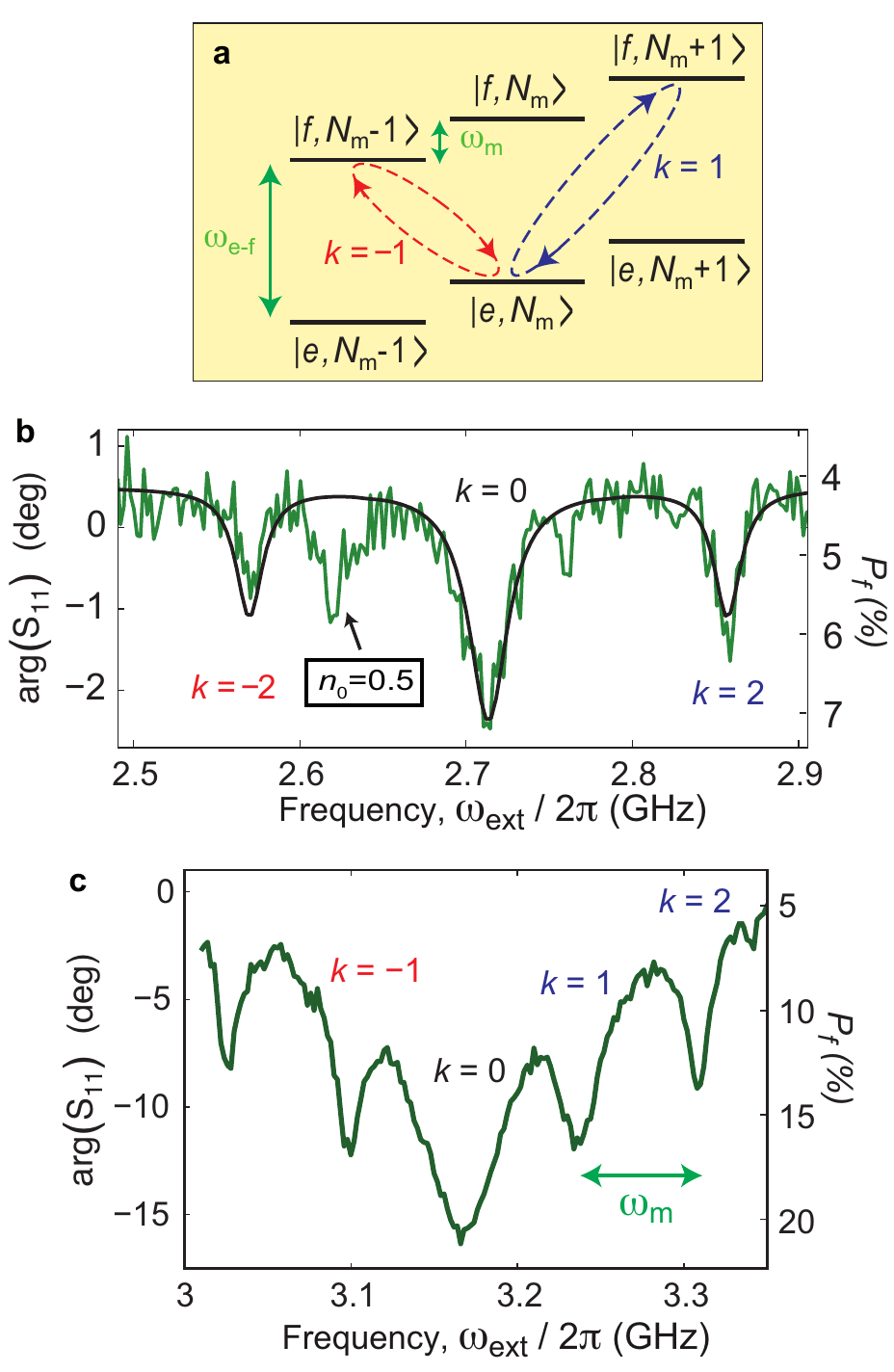}
 \caption{\textbf{Motional sideband transitions}. \textbf{a}, In a Stokes (red sideband) scattering event, the qubit exchanges (absorbs/emits) a quantum of frequency $\omega_{\m{e-f}} + k \omega_m$ with the microwave field, whereas the mechanical resonator emits/absorbs $k$ quanta. In an anti-Stokes (blue sideband) scattering event, the process is reversed. \textbf{b}, Cross-cut along the arrow from Fig.~2d at $n_x \simeq 0.6$. The solid black line is a fit from the analytical model presented in the Supplementary, Eq.~(S49), with decay rate $\gamma /2\pi = 3$ MHz, dephasing rate $\gamma_{\varphi}/2\pi = 6$ MHz, and main peak Rabi frequency $\Omega_0 /2\pi =7$ MHz. \textbf{c}, Spectroscopy with the motional gate charge corresponding to the dynamical sweet spot ($ n_x = 0.4$, see Supplementary II.H.), where the qubit is insensitive to offset charge fluctuations, $\Phi_{\m{dc}} / \Phi_0 = 0.326$. The qubit was driven hard at $\Omega_0/2\pi = 31$ MHz, making the main peak broad.}
  \label{fig:side}
\end{figure}


For investigating the energy exchange in time domain, we performed time-resolved measurements in the setup of Fig.~3c. Here, instead of a continuous microwave irradiation as in the two-tone spectroscopy of Figs.~2 and ~3, we pulsate the excitation microwave tone at varying widths, while monitoring the level $|f \rangle$ population. During the pulse, the system is expected to coherently evolve at the Rabi frequency $\Omega_k$ (for sideband $k$) between a pair of eigenstates as illustrated by arrows in Fig.~3a. Figures~4a,b show the obtained Rabi oscillations on the bare qubit transition ($k=0$) and on the first blue sideband ($k=1$). The latter indicates that the the qubit state can swap while simultaneously adding or removing one quantum in the mechanical resonator. The qubit is measured using weak continuous measurement, which here provides the population $P_f$ immediately following the pulse (supplementary, section I.C). We can map the energy exchange by varying the detuning of the excitation microwave (Fig.~4c), as well as by converting the data to frequency domain (Fig.~4d). The data shows, as expected, an increase of the Rabi frequency on detuning from a particular sideband. The coherence times are presumably limited by external high-frequency noise which causes quasiparticle dissipation and a thermal population in the qubit.

\begin{figure*}
 \includegraphics[width=0.99\linewidth]{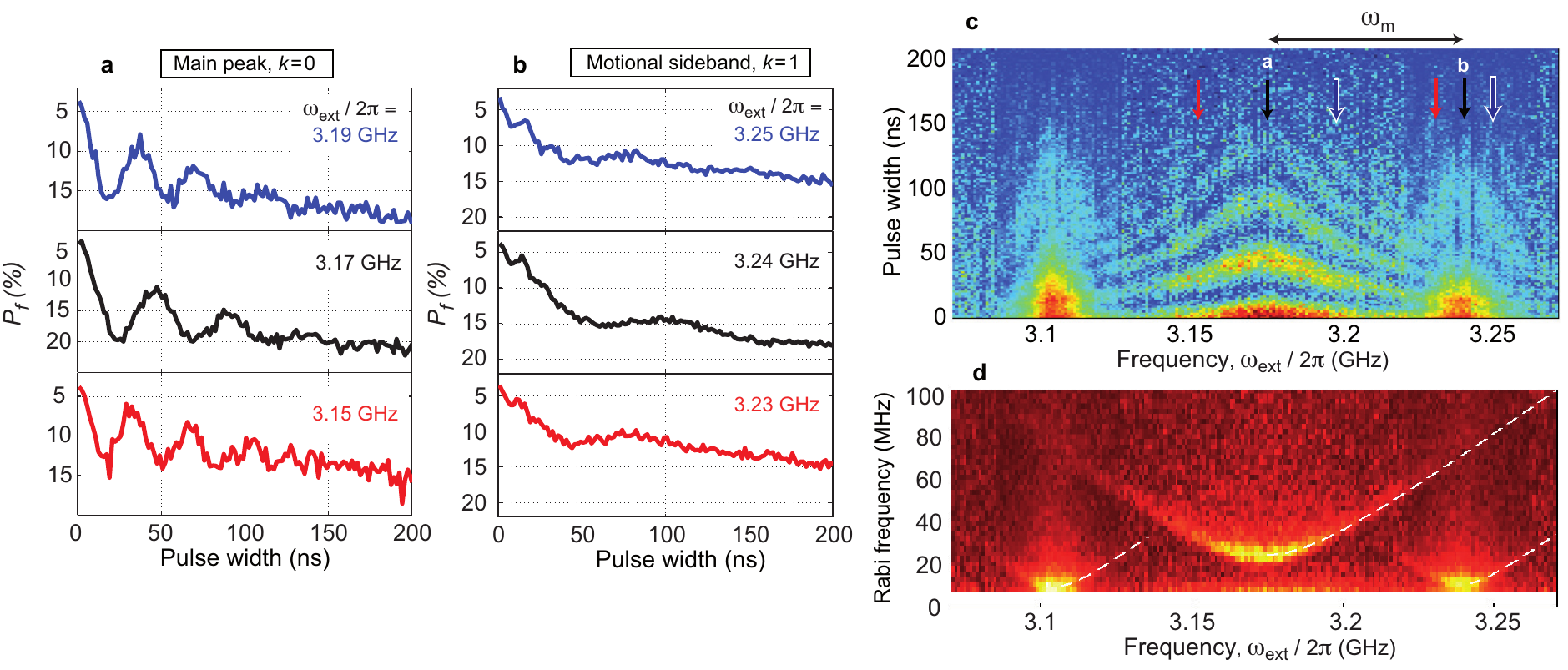}
 \caption{\textbf{Electromechanical Rabi oscillations}. \textbf{a}, Oscillations of the population of the qubit third level $|f \rangle$, measured near the bare qubit frequency $k=0$. \textbf{b}, Time-resolved evolution between phonons and qubit states, near the first blue sideband $k=1$, with the resonant Rabi frequency $\Omega_1/2\pi \simeq 9.4$ MHz. At the first minimum, around $50 ... 70$ ns, the qubit has flipped (at modest fidelity) from $|e \rangle$ to $|f \rangle$, while the mechanical resonator absorbs a quantum. \textbf{c}, Qubit population oscillations mapped as a function of the qubit excitation frequency and Rabi pulse width. An overall decay, visible in \textbf{a},\textbf{b} is subtracted for illustration clarity. The traces in \textbf{a} and \textbf{b} are along the arrows. The decay time constant, related to total decoherence rate, is $\sim 70$ ns. \textbf{d}, Fourier transform of \textbf{c} showing the Rabi frequencies of each sideband growing as a function of excitation detuning. The lines are fits to the expression $\Omega_k(\omega_{\m{ext}}) =\sqrt{\Omega_k^2 + (\omega_{\m{e-f}} - \omega_{\m{ext}} - k \omega_m)^2}$.}
  \label{fig:rabi}
\end{figure*}

The motional sideband operations facilitate the use of the full quantum toolbox developed in the field of trapped ions \cite{WinelandRMP,Wineland1996,Blatt1999} for engineering non-classical states of mechanical resonators, or long-distance entanglement of several slowly moving purely mechanical objects. At the same time, the connection of a superconducting qubit to two different resonators is a prototype of a quantum interface. The microwave resonator allows for accessing the qubits, whereas the mechanical resonator forms a building block for the conversion of quantum information between microwave light and mechanical motion. The long-lived phonons could be used as a quantum memory. While here the storage time is limited by the thermal coherence time \cite{Teufel_Nature_2011} of the mechanical resonator $\tau_T= Q_m /N_m^T \omega_m \sim 1 \, \mu$s, we note the encouraging recent findings \cite{highQBAW} with $Q_m \sim 10^{10}$ and $\tau_T$ of the order seconds. This kind of micromechanical device could further allow for quantum communication by converting phonons into flying photons using radiation pressure coupling, naturally suited for a membrane mirror. An increase of the electromechanical coupling by means of a narrower vacuum gap could bring the interaction between light and true matter all the way into the single-phonon ultrastrong regime.

\section*{Methods summary}

{\small \textbf{Experimental} \hspace{2mm} The device fabrication includes three layers of electron-beam lithography. The first lithography patterns everything else except the mechanical resonator. Aluminum is deposited by shadow evaporation at thicknesses of 20 nm and 40 nm, with an oxidation in between in order to create the Josephson tunnel junctions. The sacrificial layer separating the bridge from the transmon island is defined with PMMA used as a negative resist, under high electron dose. The mechanical resonator is defined in the third lithography. In the end, isotropic O$_2$ ashing removes the PMMA and suspends the bridge. In a tilted SEM micrograph we observe an undulating vacuum gap of about 40...100 nm between the bridge and the qubit island. 


With the electromechanical coupling set to zero by $V_{\m{dc}} = 0$, we first characterized the basic operation of the transmon-cavity circuit QED system. Based on both single-tone and two-tone spectroscopy, we determined the following qubit parameters: $E_{J1}/2\pi = 4.63$ GHz, $E_{J2} /2\pi= 6.43$ GHz, $C_{q} = 61$ fF, $E_{C} /2\pi= 318$ MHz. Here, $E_{J1}$ and $E_{J2}$ are the Josephson energies of the two junctions, and $E_J=E_{J1} + E_{J1}$.

\textbf{Theoretical} \hspace{2mm} In the qubit eigenbasis denoted by the standard Pauli matrices $\sigma_{x,y,z}$, the qubit-mechanical resonator Hamiltonian  is of the from
\begin{align}
    H \sim -\frac{\omega_{\m{e-f}}}{2} \sigma_z  + \omega_m (b^{\dagger} b + 1/2)  + g_{m,z}  ( b^{\dagger} + b) \sigma_z + g_{m,x}   ( b^{\dagger} + b) \sigma_x \,. \label{eq:sH}
\end{align}
The coupling generally contains both diagonal and transverse components $g_{m,z}$ and $g_{m,x}$, respectively. In the charge qubit limit, $g_{m,z} \approx g_m \gg g_{m,x}$, whereas in transmon limit, $g_{m,z}  \ll g_{m,x} \approx g_m$. In the present experiment, roughly $g_{m,z} \approx g_{m,x} \approx g_m$. In the detuned case, the diagonal coupling with $g_{m,z}$ is more prominent. However, we find that in the transmon limit, if we make $g_{m}/2\pi \gtrsim 25$ MHz by means of a narrower vacuum gap and/or higher dc voltage, we can obtain phonon-photon state transfer in the quantum limit within a microsecond.

}

\bibliographystyle{naturemag}
\bibliography{../HYBRIDQED2}

\textbf{Acknowledgements} We thank Matti Silveri, Erkki Thuneberg and Tero Heikkil\"a for useful discussions. This work was supported by the Academy of Finland and by the European Research Council (grant number 240387-NEMSQED) and EU-FP7-NMP-246026. The work benefited from the facilities at the Micronova Nanofabrication Center. J.-M.P. acknowledges support from V\"ais\"al\"a Foundation, Emil Aaltonen Foundation, and Kaute Foundation, and J. Li from NGSMP.

\textbf{Author contributions} M.A.S. conceived the experiment and designed the experimental set-up with P.J.H. J.L., J.-M.P, and G.S.P. designed the circuit layout. S.U.C. and J.-M.P. fabricated the samples. J.-M.P. conducted the measurements, developed theory and wrote the manuscript. S.U.C. supported the measurements. All authors commented on the manuscript.

\textbf{Additional information} The authors declare no competing financial interests. Supplementary information accompanies this paper on www.nature.com/. Reprints and permissions information is available online at http://www.nature.com/reprints. Correspondence and requests for materials should be addressed to J.-M.P.

\newpage

\renewcommand{\thefigure}{S\arabic{figure}}
\renewcommand{\theequation}{S\arabic{equation}}

\setcounter{figure}{0}
\setcounter{equation}{0}
\begin{widetext}
\Large{Hybrid circuit cavity quantum electrodynamics with a micromechanical resonator: Supplementary information}

\normalsize

\vspace{0.5cm}
J.-M.~Pirkkalainen,
S.~U.~Cho,
Jian~Li,
G.~S.~Paraoanu,
P.~J.~Hakonen,
M.~A.~Sillanp\"a\"a

\vspace{3mm}

\affiliation{O.~V.~Lounasmaa Laboratory, Low Temperature Laboratory, Aalto University, P.O. Box 15100, FI-00076 AALTO, Finland}


\section{Experimental details}
\label{sexp}

\subsection{Device fabrication}
\label{sfab}

The fabrication (Fig.~\ref{fig:sfab}) includes three layers of electron-beam lithography on a sapphire substrate. The first lithography patterns everything else except the mechanical resonator. Aluminum is deposited by shadow evaporation at thicknesses of 20 nm and 40 nm, with an oxidation in between in order to create the Josephson tunnel junctions.

The sacrificial layer separating the bridge from the transmon island is defined with PMMA used as a negative resist. We start by spin-coating 3 \% 950k PMMA, creating about 200 nm thick layer. Under high electron dose, here 9 mC/cm$^2$ at 200 pA, PMMA cross-links, shrinks in thickness down to about 40 \% and becomes insoluble. After the exposure, the non-exposed PMMA is washed away with acetone.

For the third lithography, a 800 nm thick resist serves as the deposition mask for 300 nm Al defining the mechanical resonator. After lift-off, 90 min of isotropic O$_2$ ashing at 1 Torr removes the PMMA and suspends the bridge. Although the ashing takes time, no appreciable change to qubit properties can be attributed to it.

A highly tilted SEM micrograph (Fig.~\ref{fig:sfab}b) reveals an undulating vacuum gap of about 40...100 nm between the bridge and the qubit island. Apparently, the PMMA thickness became non-uniform during the cross-linking. This makes it difficult to calculate $\partial_x C_g$ from the geometry, however, the fitted value $\partial_x C_g$ = 26 nF/m is in the proper range.


\begin{figure}[h]
 \includegraphics[width=0.7\linewidth]{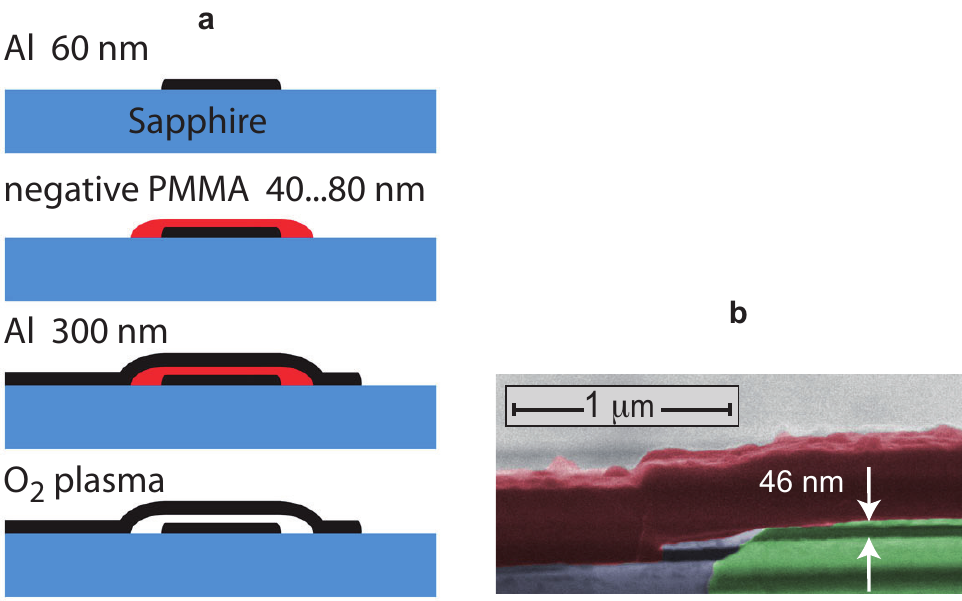}
  \caption{\textbf{Device fabrication}. \textbf{a}, Steps (from top to bottom) of the process. \textbf{b}, Gracing angle scanning electron micrograph of the mechanical resonator (red) above the superconducting qubit island (green).}
  \label{fig:sfab}
\end{figure}

\subsection{Basic characterization}
\label{sdet}

With the electromechanical coupling set to zero by $V_{\m{dc}} = 0$, we first characterized the basic operation of the transmon-cavity circuit QED system. In Fig.~\ref{fig:sspect}a we show a single-tone spectroscopy, with the avoided crossings at roughly $\Phi_{\m{dc}}  / \Phi_0 \simeq \pm 0.1$, where the qubit frequency crosses the cavity. In fact, the crossing just barely takes place, since the maximum qubit frequency is 5.0 GHz at $\Phi_{\m{dc}}  = 0$. The wings, absent in the ideal case, around the crossing points are attributed to higher transitions in the coupled system. We expect them to be due to microwave leakage via the high gate capacitance.

Two-tone spectroscopy is the main tool used in the measurements described in the main text. It is the best way to reveal the qubit spectral lines when far-detuned from the cavity. Fig.~\ref{fig:sspect}b displays such characterization over a large span of flux biases and qubit frequencies. By overlaying the results from numerical modeling, the spectral lines can be distinguished as $|g \rangle$-$|e \rangle$ and $|e \rangle$-$|f \rangle$ transitions. We adjusted the gate voltage such that $n_{0} = 0$, however, especially the $|e \rangle$-$|f \rangle$ line can be identified having split into two. The other branch ($n_{0} = 1/2$) corresponds to an extra quasiparticle on the island, excited likely by microwave leakage.

\begin{figure}[h]
 \includegraphics[width=0.99\linewidth]{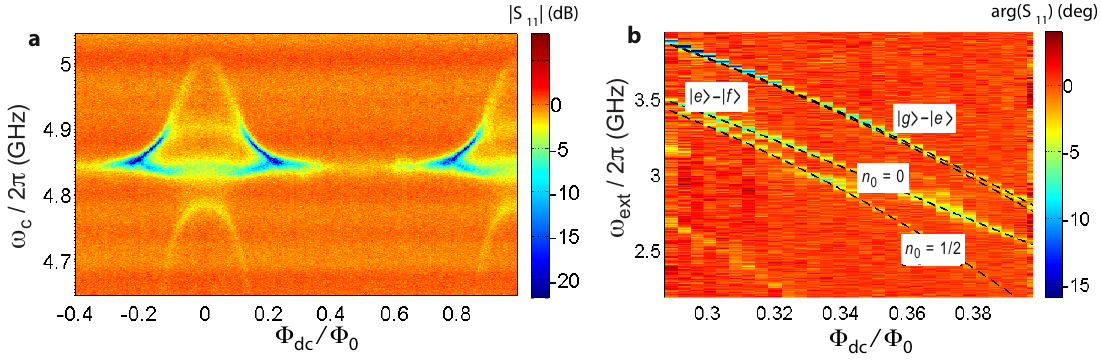}
  \caption{\textbf{Qubit spectroscopy}. \textbf{a}, One-tone spectroscopy of the cavity resonance absorption, while the effective Josephson energy is tuned by flux bias. \textbf{b}, Two-tone spectroscopy, where the probe tone $\omega_c$ is kept fixed near the bare cavity frequency (4.84 GHz), and qubit excitation frequency is swept. The dashed lines are the calculated transitions as labeled.}
  \label{fig:sspect}
\end{figure}

The data of Fig.~\ref{fig:sspect} also allows for accurate fitting of the qubit parameters; we obtain $E_{J1}/2\pi = 4.63$ GHz, $E_{J2} /2\pi= 6.43$ GHz, $C_{q} = 61$ fF, $E_{C} /2\pi= 318$ MHz, and the coupling energy to the microwave cavity $g_c/2\pi \simeq 100$ MHz.

The dc voltages applied to the mechanical resonator (acting also as a gate electrode to the qubit) make the background charge of the qubit more prone to jumps than without the dc voltage. We applied a slow digital feedback about once a minute to the gate in order to correct for the jumps. For this, we repeated a sequence consisting of measurement of the gate periodicity, followed by a small voltage offset, until the original offset was reached.

The mechanical Q-value of the present sample $Q_m \simeq 5500$ is clearly lower than typically recorded ($\gtrsim 10^5$) with aluminum beams \cite{Sulkko:2010ih} or membranes \cite{TeufelStrongG} at cryogenic temperatures. We believe this is due to either clamping losses, or the oxygen ashing changing material properties. However, the NIST design \cite{TeufelStrongG} for aluminum membrane, with $Q_m \simeq 3 \times 10^5$ could equally well integrate in the present setup, allowing for longer storage times of qubit states in the mechanical resonator.


The gate capacitance to the qubit, $C_g = 1.76$ fF, is extracted from the gate charge $n_0 = C_g V_{\m{dc}}/(2e)$ periodicity of the qubit transitions. Using the modeling of Stark shift asymmetry in Fig.~2c in main text, we can deduce the gate charge amplitude due to the direct driving. The driving voltage amplitude can then be calculated with the capacitance to the qubit. Therefore, the fitted gate charge amplitude due to the mechanical capacitance modulation translates into the derivative of the capacitance with respect to the distance, $\partial_x C_g$ = 26 nF/m. This in turn gives the coupling between the qubit and the mechanical resonator to be $g_m/2 \pi$ = 9 MHz with $V_{\m{dc}}$ = 10 V which was the maximum voltage applied to the device. The obtained values of the driving voltage agree well with the estimated attenuation of the drive cable, although the latter has inaccuracies of several dB.

\subsection{Qubit populations}
\label{spop}

According to the standard circuit QED recipe, the qubit state can be extracted from the measured phase shift $ \m{arg}(S_{11})$ of the probe tone. The cavity frequency pull by the qubit (for the qubit lowest transition) is $\Delta \omega_c = \pm g_c^2/(\omega_c - \omega_{\m{g-e}})$, depending on the qubit state. This gives rise to the phase shift $ \Theta_{\m{g,e}} \equiv  \m{arg}(S_{11}) = \pm 2 g_c^2/\gamma_E(\omega_c - \omega_{\m{g-e}})$. Any pair of higher qubit transitions has a distinct $g_c$, as well as different detuning from the cavity. Hence they give rise to a unique phase shift, and consequently in principle also the higher states can be identified. However, the procedure becomes progressively involved the more qubit states participate in the dynamics.

During the two-tone spectroscopy or Rabi oscillation measurements, we suppose that the populations $P_g, P_e, P_f$ of the three lowest qubit states, respectively, are affected by spontaneous relaxation, excitation and relaxation by noise, and intentional driving. We will adopt a notation where the transition rate from level $i$ to $j$ is marked by the subscript i-j. The spontaneous relaxation rates are $\gamma_{\m{e-g}}$ and $\gamma_{\m{f-e}}$ (the latter is called $\gamma$ for simplicity in main text), and the transition rates due to external noise are $\Gamma_{\m{e-g}} = \Gamma_{\m{g-e}}$ and $\Gamma_{\m{e-f}} = \Gamma_{\m{f-e}}$. The excitation tone can be set on resonance to either $\omega_{\m{g-e}}$ or $\omega_{\m{e-f}}$, with the transition rates $\Omega_{\m{g-e}}$ and $\Omega_{\m{e-f}}$ (the latter equals the Rabi frequency $\Omega_0$).

We write down the basic rate equations for the populations. As an example, for the $e$ level it reads when the excitation tone is on resonance to $\omega_{\m{e-f}}$, 
\begin{equation}
\dot{P_e} = \Gamma_{\m{g-e}} (P_g - P_e) - (\Omega_{\m{e-f}} + \Gamma_{\m{e-f}}) P_e + (\Omega_{\m{e-f}} + \Gamma_{\m{e-f}}) P_f - \gamma_{\m{e-g}} P_e + \gamma_{\m{f-e}} P_f.
\end{equation}
The condition of steady-state allows for solving the populations in three cases: when the excitation tone is off-resonance (a situation in which case it can be ignored), or when it is on-resonance to either transition. These three cases correspond to the red background, or the $g$-$e$ or $e$-$f$ transition lines in Fig.~\ref{fig:sspect}b, respectively. The measured phase shift is in either case $\m{arg}(S_{11}) = P_g \Theta_{\m{g}} + P_e \Theta_{\m{e}} + P_f \Theta_{\m{f}}$.

Using $H = H_q + H_c +H_{qc}$ (see Eqs.~(\ref{eq:transcavHam2},\ref{eq:transcavHams})) we calculate numerically $\Delta \omega_c$ for the three qubit states, and then $\Theta_{\m{i}}$ from the measured cavity phase response. With the flux biases used in the data in main text, we obtain $\Theta_{\m{e}} -\Theta_{\m{g}} \sim -50^{\m{o}}$,  $\Theta_{\m{f}} -\Theta_{\m{g}}\sim -100^{\m{o}}$. Studying the qubit transition width as a function of excitation power allows for deducing the $\Omega$'s as well as total decoherence rate. $\gamma_{\m{f-e}}$ is roughly obtained from the time-domain measurements in Fig.~4 in the main text. We also suppose $\gamma_{\m{e-g}} = \gamma_{\m{f-e}}$, but we still have to figure out the noise-induced rates. However, we obtain from the analysis that insensitive to these, one can reliably obtain the populations based on the absolute and relative depths of the $g$-$e$ or $e$-$f$ dips with each other. For example, a fully saturated $e$-$f$ transition as in Fig.~3c in main text, having $\m{arg}(S_{11}) \sim -17^{\m{o}}$, has $P_f \sim 20$ \%.

In the pulsed Rabi oscillation measurement, the state can be measured either by a measurement pulse that follows the Rabi pulse, or by weak continuous readout \cite{WallraffPRL05}. In all data shown, we used the latter. The Rabi pulse is repeated at the repetition frequency $\gamma_{\m{prf}}$. We use the previous calibration from phase shift on the $e$-$f$ transition into $P_f$. A rigorous treatment would consider a full time-dependent three-level Lindblad master equation. However, since we are not attempting to prove high fidelities, a rough model is sufficient for the moment. Here one also needs to consider the fact that the signal decays on the $\gamma_{\m{f-e}}$ rate after the Rabi pulse \cite{WallraffPRL05}. The average signal level in this measurement is thus multiplied by a factor of $\gamma_{\m{prf}}/\gamma_{\m{f-e}} \LL[ 1-\exp(-\gamma_{\m{f-e}}/\gamma_{\m{prf}}) \RR] $ as compared to continuous-wave spectroscopy.

\subsection{Quasiparticle decoherence}
\label{scoherence}

We believe the somewhat low coherence of the qubit is due to issues with shielding of the experimental space, e.g., the sample holder was not radiation tight. In recent literature, dissipation due to tunneling of quasiparticles has been associated to loss of coherence in superconducting qubits \cite{IBMqp2011,MartinisQP2011,DevoretQP2011}. Experiments often report effective quasiparticle temperatures in the range of 150 mK \cite{DelsingQP08,AumentadoQP09,Paik_PRL_2011,IBMqp2011}, although the qubit is found below 50 mK.

\begin{figure}[ht!]
 \includegraphics[width=0.5\linewidth]{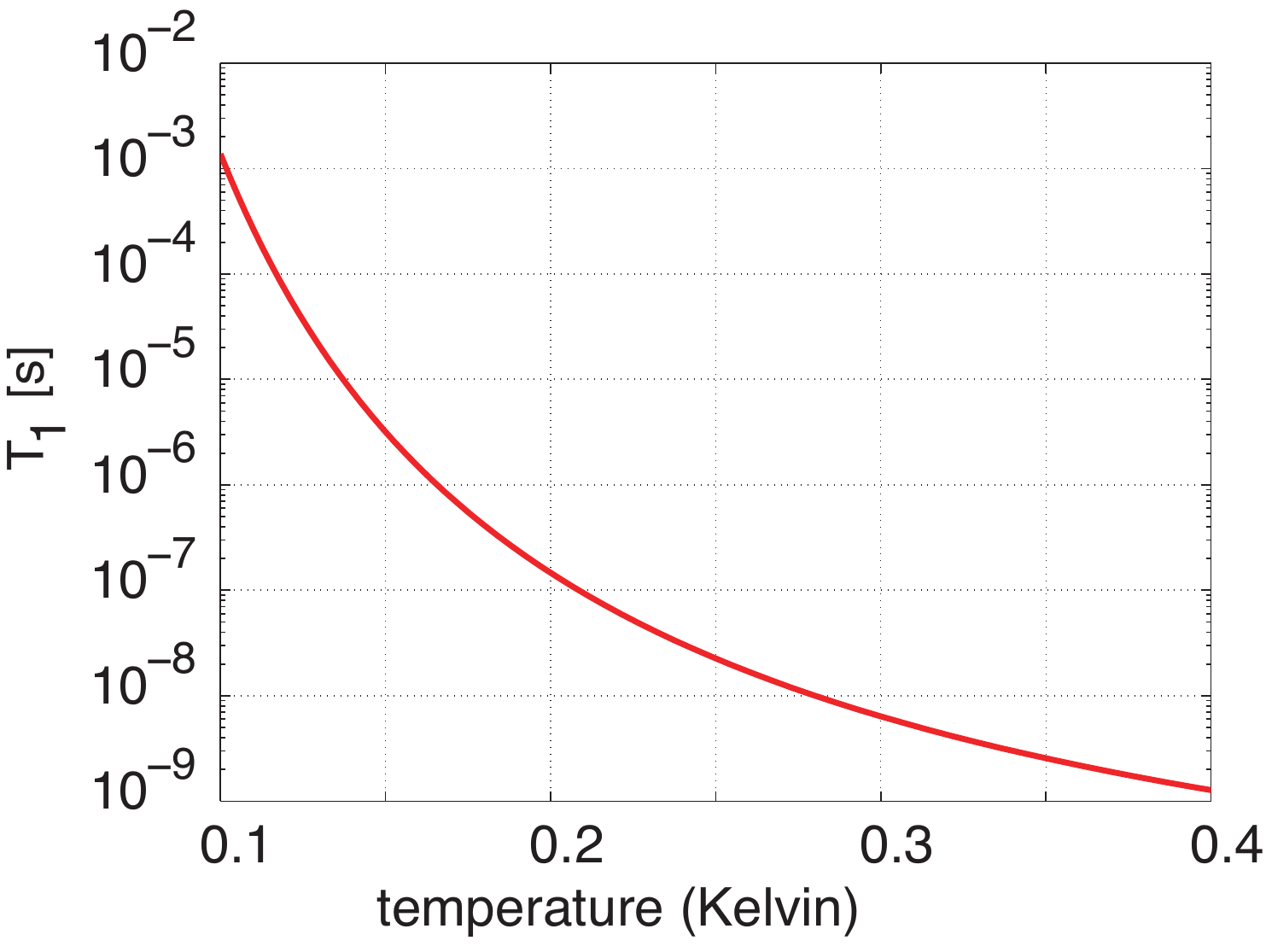}
 \caption{\textbf{Quasiparticle relaxation in transmon}. We apply results for the split transmon, Eq.~(126) from Ref.~\cite{DevoretQP2011}. The plot shows the qubit relaxation rate due to these quasiparticles with the experimental parameters as a function of the equilibrium temperature of the environment.}
  \label{fig:sqprelax}
\end{figure}



The source of the quasiparticles is not clear, but infrared radiation from high-temperature parts of the apparatus has been considered a likely reason. Based on level occupations in the absence of drive, our qubit is at a temperature of 140 mK. We can apply the results from Ref.~\cite{DevoretQP2011} for quasiparticle density and the relaxation rate in equilibrium with the environment, see Fig.~S3. In order to obtain the measured range $2\pi /\gamma \sim 60$ ns, we would need a temperature $\sim 200 ... 250$ mK. Although this is somewhat higher than the measured temperature of the qubit, one has to bear in mind that the situation is highly out of thermal equilibrium, and the analysis may not be applicable as such. A quasiparticle temperature higher than the qubit temperature is also supported by the typical numbers mentioned above.
 
An encouraging example for improving shielding is provided by the IBM group \cite{IBMqp2011}, who by a simple improvement of shielding using absorptive material, the same time got rid of both unexplained qubit thermal excitation and a compromised coherence time.

\vspace{0.5cm}

\subsection{Mechanical Stark shift}
\label{sexpstark}

Next, we discuss additional data on the mechanical Stark shift.  In figure~\ref{fig:stark025}, the gate charge is tuned to $n_{0} = 0.25$, at which point the qubit transition frequency is insensitive to quasiparticle jumps and only one transition line is observed. Equation (2) in main text, indeed, gives no dependence of the transition frequency on the modulation amplitude. Accurate numerics, plotted as a dashed line, however, displays some dependence owing to deviations from simple sinusoidal charge dispersion. Due to strong $1/f$ charge noise when off the gate charge sweet spot ($n_0 = 0; 0.5$), the transition peak as well as the phonon sidebands are nearly absent except at the dynamical sweet spots (see Fig.~\ref{fig:ChargingEnergy}b).

\begin{figure}[h]
 \includegraphics[width=0.5\linewidth]{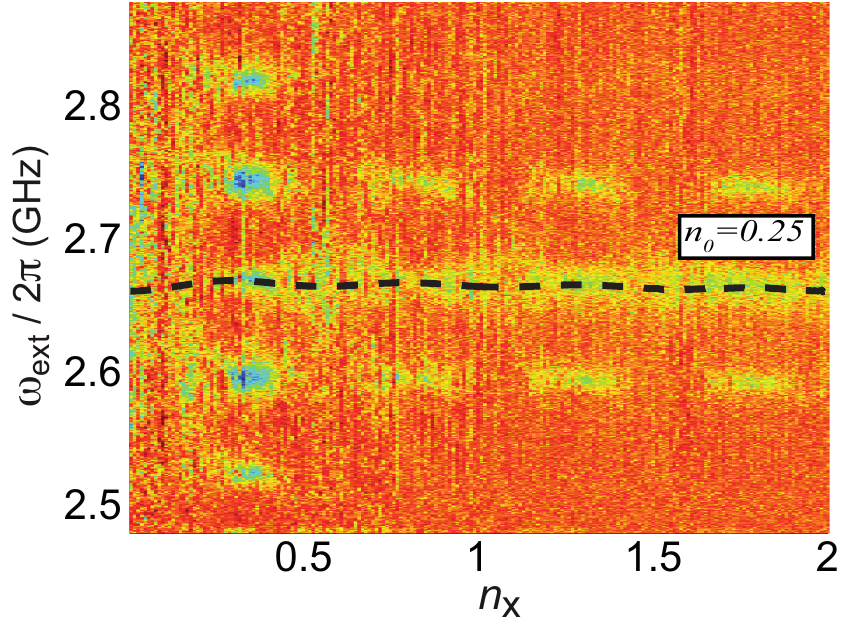}
  \caption{\textbf{Mechanical Stark shift}. Similar to Fig.~2d in main text, but the gate offset is set to the maximum slope, $n_0 = 0.25$, of the modulation curve. The measured population of $|f \rangle$ is proportional to the color code. At this gate offset, the Stark shift is minimal as expected from both analytics and numerics. The dashed line is from full numerical Floquet modeling.}
  \label{fig:stark025}
\end{figure}

\section{Theoretical modeling}
\label{stheory}

\subsection{Hamiltonian}
\label{shamilt}

We are considering the tripartite system consisting of the transmon qubit, microwave cavity, and the mechanical resonator, as displayed in Fig.~\ref{fig:TransmonNEMS}. The displacement $x$ of the mechanical resonator is defined as the maximum deflection amplitude at halfway between clamping points of the beam. This definition is accompanied by the effective mass $m$ of the mode in question, the lowest flexural mode, being $3/8$ times the total mass of the resonator.

The displacement can in the following be either a classical variable or an operator. In the latter case, we have
\begin{equation}
x =  x_{\m{zp}} \LL( b^{\dagger} + b \RR)\,.
\end{equation}
The zero-point oscillation rms amplitude is
$$
x_{\m{zp}} = \sqrt{\frac{\hbar}{2 m \omega_m}} \,.
$$

The qubit is connected to the mechanical resonator via the gate capacitance $C_g(x)$. The displacement gives rise to a motional gate charge defined as
\begin{equation}
\label{eq:nxdef}
n_x  = \frac{d C_g(x) }{d x} \frac{V_{\m{dc}} }{2e }x  \,.
\end{equation}

The transmon qubit \cite{transmon} is schematically the same as the split Cooper-pair box, consisting of a superconducting loop interrupted by two Josephson junctions with Josephson energies $E_{J1}$ and $E_{J2}$. For future use, we define the total Josephson energy $E_J$, and the difference $E_{J-}$. They are tunable by the flux $\Phi_{\m{dc}}$ through the superconducting loop:
\begin{equation}
\begin{split}
E_{J} \LL( \Phi_{\m{dc}} \RR)=&  \LL(E_{J1}+E_{J2} \RR) \cos \LL(\pi \frac{\Phi_{\m{dc}}}{\Phi_0}\RR)  \\ 
    E_{J-} \LL( \Phi_{\m{dc}} \RR)= &\LL(E_{J1}-E_{J2} \RR) \sin \LL( \pi \frac{\Phi_{\m{dc}}}{\Phi_0}\RR) \,. 
    \end{split}
\end{equation}
We linearize the dependence of the capacitance on displacement:
\begin{equation}\label{eq:dCdx}
\begin{split}
C_g(x) = C_g + \frac{d C_g(x)}{dx}x =C_g \LL( 1 + \frac{C_g'}{C_g} x\RR) \,,
    \end{split}
\end{equation}
where we denoted the undisplaced gate capacitance by $C_g(x=0) \equiv C_g$. The capacitance in parallel with the junctions is denoted by $C_q$, the sum of the capacitances of the individual junctions and of an interdigital capacitor. The total capacitance of the qubit is the sum of $C_q$, $C_g(x)$, and of $C_t$ which is the coupling capacitor to cavity. The values are supposed to be related as $C_g, C_t \ll C_q$. The (single-electron) charging energy is
$$ E_C = \frac{e^2}{2 \LL[ C_q + C_g(x) + C_t\RR] } \simeq  \frac{e^2}{2 C_q } \, .$$

\begin{figure}[!h]
    \includegraphics[width=10cm]{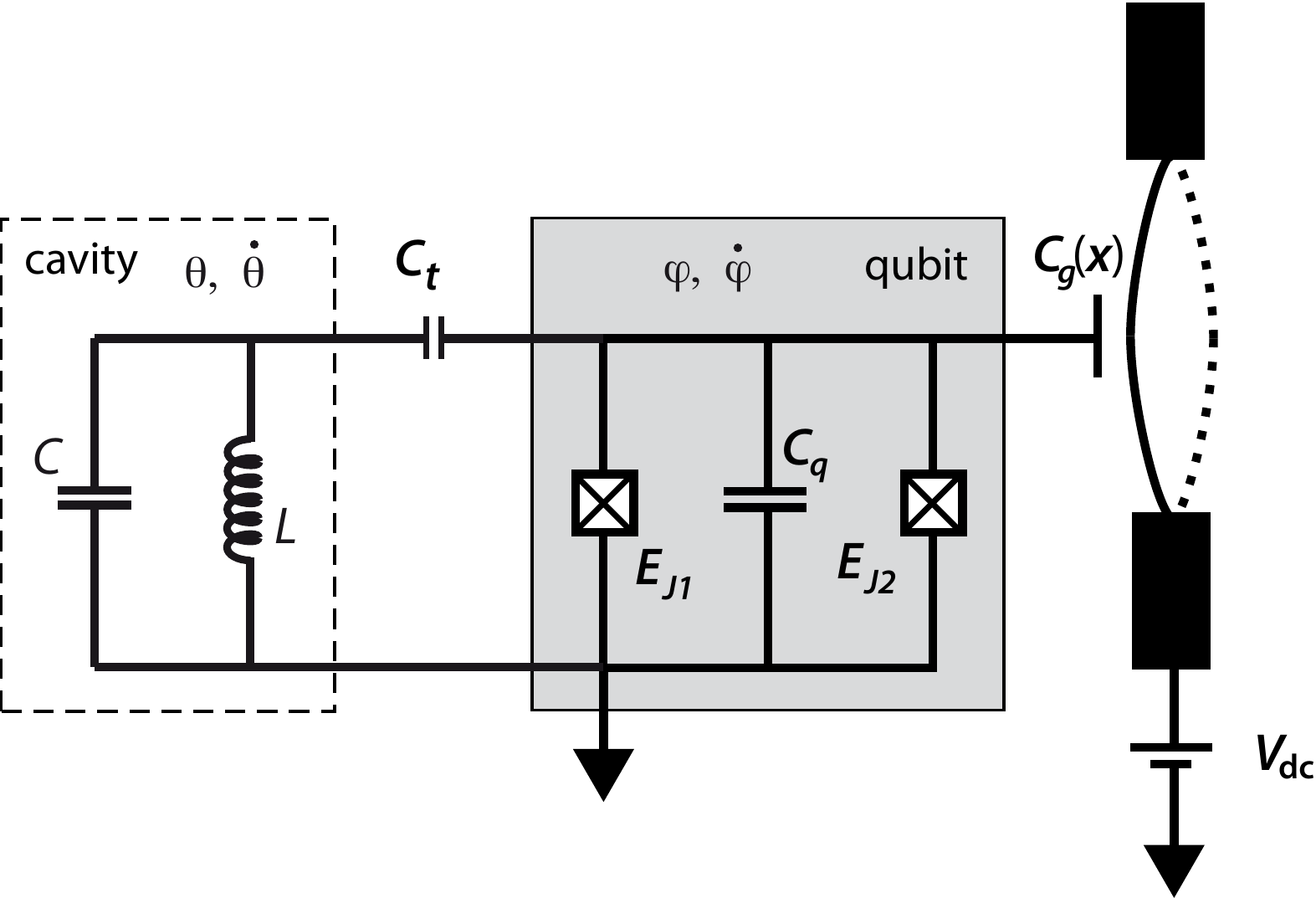}
    \caption{Circuit schematics of a transmon qubit (shaded box) coupled capacitively to a micromechanical resonator via a gate capacitance $C_g(x)$, as well as to a microwave cavity represented as an $LC$ resonator (dashed box).} \label{fig:TransmonNEMS}
\end{figure}

Let us derive the Hamiltonian for the entire tripartite system. The canonical coordinates $\varphi$ and $\theta$ denote the phases of the qubit island, and of the cavity, respectively. The qubit island charge $Q$ is related to the number of Cooper pairs $n$ on the island as
$$
Q = 2en = C_g ( \dot{\varphi} - V_{\m{dc}}) + C_t( \dot{\varphi} - \dot{\theta} ) + C_q \dot{\varphi} \,.
$$
One writes down the Lagrangian which is a function of the generalized coordinates and their time derivatives. We use the subscript $q$ for the qubit, $c$ for the cavity, and $m$ for the mechanical resonator. The Hamiltonian will consist of the Hamiltonians of the qubit $H_q$, the cavity $H_c$, the mechanical resonator $H_m$, and of their couplings. We expand in the small parameter $\frac{C_g'}{C_g}x$, and obtain
\begin{equation}\label{eq:transcavHam2}
\begin{split}
 H = & H_q + H_c + H_m +H_{qc} + H_{qm} +H_{cm} +H_{qcm}  +H_{c2m} \,.
   \end{split}
\end{equation}
The terms are
\begin{equation}\label{eq:transcavHams}
\begin{split}
    &H_q = 4 E_C \LL(n - n_0 \RR)^2 - E_{J}  \cos \LL( \varphi \RR) -E_{J-}   \sin \LL( \varphi \RR) \\
    &H_c =  \hbar \omega_c \LL(a^{\dag}a +\frac{1}{2}\RR) \\
    &H_m =  \hbar \omega_m \LL(b^{\dag}b + \frac{1}{2}\RR)  \\
    &H_{qc} =  \hbar g_{c} (n_0-n) (a^{\dag}+a) \\
    &H_{qm} =  \hbar g_{m} (n_0-n) (b^{\dag}+b) \\
  &H_{cm} =  \hbar g_{cm} (a^{\dag}+a) (b^{\dag}+b) \\
    &H_{qcm} =  \hbar g_{qcm} n (a^{\dag}+a) (b^{\dag}+b) \\
    &H_{c2m} = \hbar g_{c2m} (a^{\dag}+a)^2 (b^{\dag}+b) \,,
      \end{split}
\end{equation}
and the coupling energies read
\begin{equation}\label{eq:transcavHamsg}
\begin{split}
   &\hbar g_{c} = \frac{C_t e}{ C_q} \sqrt{\frac{2 \hbar \omega_c}{C}} \\
   &\hbar g_{m} = x_{\m{zp}}  V_{\m{dc}} C_g' \frac{2e}{C_{q}} \\
   & \hbar g_{cm} =  \frac{x_{\m{zp}} C_g'  C_t V_{\m{dc}}}{2C_q} \sqrt{\frac{2 \hbar \omega_c}{C}} \\
   & \hbar g_{qcm} =  \frac{x_{\m{zp}}  C_g' C_t e}{C_q^2} \sqrt{\frac{2 \hbar \omega_c}{C}} \\
   & \hbar g_{c2m} =   \frac{ C_t^2}{C_q^2}\frac{x_{\m{zp}}  C_g' \hbar \omega_c}{2 C}  \,.
      \end{split}
\end{equation}

The direct cavity-mechanical resonator linear interaction $H_{cm}$ is usually negligible, owing to different energy scales, $\omega_m \ll \omega_c$. The last term $H_{c2m}$ in Eqs.~(\ref{eq:transcavHams}) and~(\ref{eq:transcavHamsg}) is a radiation-pressure interaction from phonons to cavity. Although generally small, it can provide a medium for finding and characterization of the mechanical resonance via the "circuit optomechanics" methods \cite{Regal:2008di} by pumping the cavity up to photon numbers $\gg 1$. The second-last term $H_{qcm} $ also becomes in the dispersive limit approximately a radiation-pressure term \cite{ArmourNJP}, however, it becomes irrelevant since a high photon number destroys the qubit operation.

The relevant physics discussed here is thus due to the Hamiltonian of Eq.~(1) in the main text,
\begin{equation}\label{eq:transcavHam3}
\begin{split}
 H =  H_q + H_c + H_m +H_{qc} + H_{qm}  \,,
   \end{split}
\end{equation}
which describes the qubit coupled to two harmonic oscillators.

%



\subsection{Quantum treatment of the electromechanical coupling}
\label{squantum}

The qubit-phonon interaction part in Eq.~(1) in the main text is (setting $\hbar=1$)
\begin{align}\label{eq:JC0}
    H=  H_{\mathrm{q}}  + \omega_m (b^{\dagger} b + 1/2)  + g_m ( n_{0} -n) ( b^{\dagger} + b). 
\end{align}
One can diagonalize the qubit, and rotate the coupling term into the qubit eigenbasis. In the following, we discuss the $|e \rangle$-$|f \rangle$ transition unless explicitly mentioned. Generally, the coupling will have both diagonal and transverse components:
\begin{align}
    H \sim -\frac{\omega_{\m{e-f}}}{2} \sigma_z  + \omega_m (b^{\dagger} b + 1/2)  + g_{m,z}  ( b^{\dagger} + b) \sigma_z + g_{m,x}   ( b^{\dagger} + b) \sigma_x \,. \label{eq:sH}
\end{align}

As seen in Fig.~\ref{fig:qHamilt} where we plot the coupling energies $g_{m,z}$ and $g_{m,x}$, the diagonal coupling is usually quite small (apart from the lowest qubit frequencies around $\Phi_{\m{dc}}  / \Phi_0 \simeq  0.5$), and that $g_{m,x} \sim g_m$. Therefore, Eq.~(\ref{eq:sH}) becomes the Jaynes-Cummings Hamiltonian. We note that this holds in the low-$N_m$ limit, see below in this section.

\begin{figure}[!h]
    \includegraphics[width=16cm]{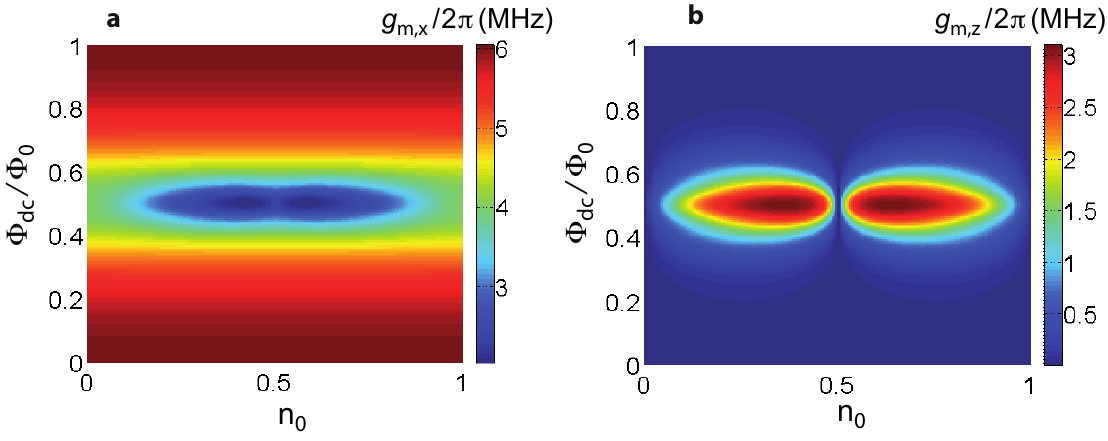}
    \caption{\textbf{Qubit-mechanical resonator Hamiltonian in qubit eigenbasis}. We have  used parameters of our experiment. \textbf{a}, transverse coupling energy for $|e \rangle$-$|f \rangle$ transition. \textbf{b}, corresponding diagonal coupling energy.} \label{fig:qHamilt}
\end{figure}

The eigenstates from Eq.~(\ref{eq:sH}) are the dressed states of the qubit and of the harmonic oscillator Fock states $| N_m \rangle$ with $N_m$ phonons:
\begin{equation}\label{eq:JCvect}
\begin{split}
|+,N_m \rangle =& \cos \left( \frac{\Theta_N}{2} \right) |f,N_m
\rangle + \sin \left( \frac{\Theta_N}{2} \right) | e, N_m+1 \rangle \\
|-,N_m \rangle =& - \sin \left( \frac{\Theta_N}{2} \right) |f,N_m
\rangle + \cos\left( \frac{\Theta_N}{2} \right) | e, N_m+1 \rangle \,.
\end{split}
\end{equation}
Here,
\begin{equation}\label{eq:tanThetaJC}
    \tan \Theta_N = \frac{2g_m\sqrt{N_m+1}}{\Delta} \, , \: \: \: \: \: \: \: \:
    \Delta = \omega_{\m{e-f}} -  \omega_m \,.
\end{equation}
The eigenenergies are 
\begin{equation}\label{eq:JClevEn}
    E_{\pm,N} =  \omega_m (N_m+1) \pm \frac{1}{2}\sqrt{4g_m^2(N_m+1)+\Delta^2} \, .
\end{equation}

We can diagonalize Eq.~(\ref{eq:JC0}) numerically, and obtain results which are close to Eqs.~(\ref{eq:JCvect}) - (\ref{eq:JClevEn}). In Fig.~\ref{fig:qstark} we show the Stark shifts when the number of phonons increases from zero to one,  which are quite small due to large detuning. For the higher transitions it is $\omega_{\m{e-f}}(N_m = 1) - \omega_{\m{e-f}}(N_m = 0)$. The signs are opposite due to a half gate period offset. See a discussion on observability in this regime in section \ref{slown}.

\begin{figure}[!h]
    \includegraphics[width=14cm]{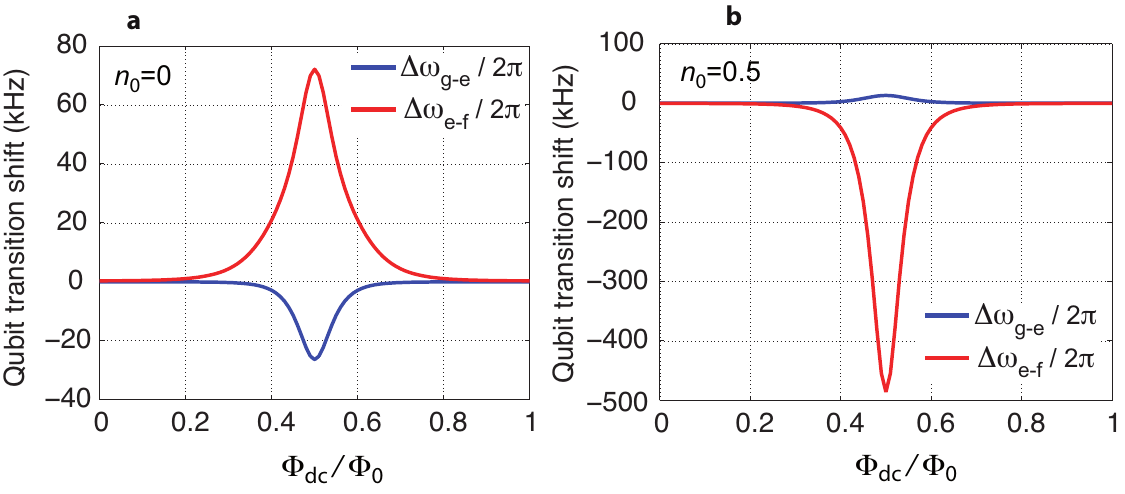}
    \caption{\textbf{Stark shift of the qubit frequency in the quantum limit}. The phonon number is changed from $N_m = 0$ to $N_m = 1$. We have used parameters of the present experiment, and shifts for the $|g \rangle$-$|e \rangle$ and $|e \rangle$-$|f \rangle$ transitions are plotted. \textbf{a}, $n_0 = 0$. \textbf{b}, $n_0 = 0.5$. Here, the pronounced effect for $|e \rangle$-$|f \rangle$ line is due to the proximity of the transition from $|f \rangle$ to the fourth qubit level, to the mechanical frequency.} \label{fig:qstark}
\end{figure}

The degree of mixing is given by the angle $\Theta_N$ which basically depends on the ratio of coupling to the qubit frequency, and on the number of phonons. By increasing the phonon number, therefore, the eigenstates become substantially dressed. Another way to indicate the increased coupling with higher Fock states is that from the Jaynes-Cummings results in Eq.~(\ref{eq:tanThetaJC}),~(\ref{eq:JClevEn}) one sees that an effective coupling is
$$
g^* = g_m\sqrt{N_m+1} \,,
$$
which in the present case, given a maximum about $N_m \sim 2 \times 10^6$, would exceed the qubit frequency.

The above results assume the validity of the linearized interaction Eq.~(\ref{eq:sH}), however, this assumption does not hold any more when the motional gate charge becomes comparable to about $n_x \gtrsim 0.1$. A rough estimate for the mixing angle is then given by using for the effective coupling the amplitude for the transmon charge dispersion, see below Eq.~(\ref{eq:eps}), \emph{viz.},~$g^* \sim \epsilon_n \sim (2\pi) \cdot 200$ MHz. 

A more accurate method to find the eigenstates beyond the linear regime in the high $N_m$ case is diagonalization of the Floquet matrix (section \ref{sfloq}). Here, the coupling of the qubit to the mechanical resonator is treated semiclassically, \emph{i.e.}, by considering the latter as a classical field coupling to the qubit: 
\begin{equation}\label{eq:nx}
n_g(t) = n_{0} + n_x \cos(\omega_g t) \,.
\end{equation}
The general expression for the state vector, written explicitly for the terms containing three lowest levels of the qubit, is:
\begin{equation}\label{eq:floqvec}
|\Psi \rangle = \sum_{N_m} G(N_m) | g, N_m \rangle + E(N_m) | e, N_m \rangle + F(N_m) | f, N_m \rangle + ...\,.
\end{equation}
In Fig.~\ref{fig:floqvec} we plot the absolute values of expansion coefficients $G(N_m)$, $E(N_m)$, $F(N_m)$, as an example, to the state vector which corresponds to the qubit excited state $ | e \rangle$. Qualitatively similar results are obtained for the other qubit states.

\begin{figure}[!h]
    \includegraphics[width=17cm]{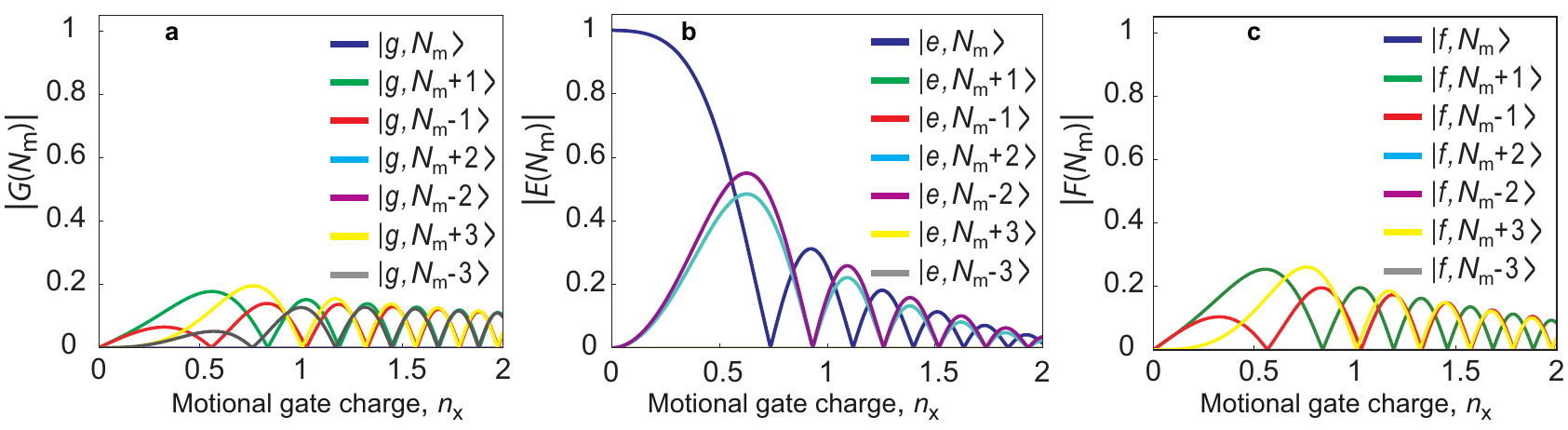}
    \caption{\textbf{Decomposition of qubit-mechanical resonator eigenstates}. We have used parameters of the present experiment, more precisely, those pertaining to the sideband Rabi experiment, Figs.~3c and 4 in main text, and $n_0 = 0$. We plot  Eq.~(\ref{eq:floqvec}) calculated from the Floquet solution, for the state which is in correspondence to the state $ | e \rangle$. } \label{fig:floqvec}
\end{figure}

We then obtain that the dressing of the eigenstates at the sideband Rabi oscillation case $n_x \simeq 0.4$ remains quite moderate, at the level of $\sim 20$ \% in the experiment. In the first approximation, the eigenstates in this case are hence states with the qubit in either $ | e \rangle$ or $ | f \rangle$, and the mechanical resonator with a certain $N_m$ number of quanta. In the deep nonlinear regime, however, the eigenstates become complicated.

The benefit of using the higher $ | e \rangle$-$ | f \rangle$ state manifold instead of $ | g \rangle$-$ | e \rangle$ is the higher coupling obtained with the former, but the situation is not straightforward. The transverse coupling $g_{m,x}$ for the lowest transition is smaller roughly by $\sqrt{2}$ from that in Fig.~\ref{fig:qHamilt}. The diagonal coupling $g_{m,z}$, however, is an order of magnitude smaller. The most prominent effect here is that in the deep nonlinear regime, showing up in the 10-fold difference in the amplitude of charge dispersion, Eq.~(\ref{eq:wntransmon}).

\subsection{Charge qubit approximation}
\label{sstark}

In the rest of the Supplementary, the mechanical motion is treated as a classical field coupling to the qubit, i.e., Eq.~(\ref{eq:nx}). In the basis of number states, the general Hamiltonian of the split Cooper-pair box, or transmon, reads
\begin{equation}\label{eq:cptHamiltn3}
\begin{split}
    H_q = & \sum_n \Big\{ 4 E_C \LL(n - n_g \RR)^2  | n\rangle 
   -\frac{1}{2}E_{J} \LL( | n + 1 \rangle +  | n - 1 \rangle \RR)-\frac{i}{2}E_{J-} \LL( | n + 1 \rangle - | n - 1 \rangle \RR) \Big\} \langle n |  \,.
    \end{split}
\end{equation}

Let us first consider the case of a charge qubit, which has $E_C \gtrsim E_J$. Although the transmon qubit operates in the extreme limit $E_C \ll E_J$, the present discussion provides, nonetheless, useful insight. For the moment, we ignore the asymmetry of the Josephson energies. 

Restricting to a subspace involving two states, Eq.~(\ref{eq:cptHamiltn3}) can be written using the Pauli matrices. With the two lowest-energy states $n=0, -1$, as well as $n_g$ between $0$ and $-1/2$, we obtain
\begin{equation}\label{eq:H01}
\begin{split}
    H_{01} = -2 E_C (1 + 2n_g) \sigma_z -\frac{1}{2}E_J \sigma_x \,.
    \end{split}
\end{equation}
The Hamiltonian $   H_{01}$ has the eigenstates $|g \rangle$ and  $|e \rangle$, the ground state and the first excited state, respectively. Of most interest for the present work is the next higher transition:
\begin{equation}\label{eq:H12}
    H_{12} = -8 E_C n_g \sigma_z -\frac{1}{2}E_J \sigma_x \,,
\end{equation}
which has the eigenstates $|e \rangle$ and  $|f \rangle$. Hence, the effective charging energy, and thus the charge dispersion of the energy, is enhanced by a factor of two by considering the first excited and second excited charge states. Generally, the bare charging energy is multiplied by a factor $k$ for a transition between levels $k-1$, $k$ in the charge qubit model. Notice that the gate charge periodicity in Eq.~(\ref{eq:H12}) is offset by half a period with respect to the lowest transition, Eq.~(\ref{eq:H01}).

We swap the $z$ and $x$ spin indices in Eq.~(\ref{eq:H01}), for the purpose of following the standard that the spin is pointing along the $z$ axis. The coupling of the driving field via charging energy is then perpendicular to the pseudospin direction. The two-charge state Hamiltonian describing the ground and first excited states $g$ and $e$ is:
\begin{equation}\label{eq:H01b}
\begin{split}
    H_{01} =  -\frac{1}{2}E_J \sigma_z -2 E_C (1 + 2n_g) \sigma_x \,,
    \end{split}
\end{equation}
with $n_g$ between  $0$ and $-1/2$, see Fig.~\ref{fig:ChargingEnergy}a. Beyond these values setting the limits of the linear regime (in the charge qubit case), the charging energy is periodic. We can approximate this dependence as sinusoidal, such that Eq.~(\ref{eq:H01b}) becomes
\begin{equation}\label{eq:H01bsin}
    H_{01} \simeq -\frac{1}{2}E_J \sigma_z  - E_c  \LL[1 + \cos(2 \pi n_g)  \RR] \sigma_x \,.
\end{equation}
%

Using Eq.~(\ref{eq:nx}), we obtain all harmonics by the Bessel expansion of the sinusoid of a sinusoid,
%
%
while the dominant term is time-independent:
\begin{equation}\label{eq:dc}
    H_{01} \simeq -\frac{1}{2}E_J \sigma_z  - E_c   \Big\{1 + \cos(2 \pi n_{0}) J_0(2 \pi n_x) \Big\}  \sigma_x  .
\end{equation}
The difference between level spacings of Eq.~(\ref{eq:dc}) and that of the uncoupled qubit, Eq.~(\ref{eq:H01bsin}) with $n_g(t) = n_{0}$, yields the Stark shift for the transition between the ground state $g$ and the first excited state $e$:
\begin{equation}\label{eq:acstark1}
    \Delta \omega_{g-e} 
   \simeq \frac{2 E_C^2 }{E_J} \Big[2 \cos(2 \pi n_{0}) (J_0(2 \pi n_x) - 1)
    +\cos^2 (2 \pi n_{0}) (J_0^2(2 \pi n_x)  -1)\Big].
\end{equation}
Similar to Eq.~(\ref{eq:H01bsin}), we can approximate the higher transition Eq.~(\ref{eq:H12})
\begin{equation}\label{eq:hdriven12}
    H_{12} \simeq -\frac{ E_J}{2} \sigma_z - 2 E_c  \LL[ 1-\cos(2 \pi n_g) \RR] \sigma_x
\end{equation}
and we obtain a Stark shift for the next higher pair of states $e$ and $f$, enhanced by a factor of 4 over Eq.~(\ref{eq:acstark1}):
\begin{equation}\label{eq:acstark12}
    \Delta \omega_{e-f} 
    \simeq \frac{8 E_C^2 }{E_J} \Big[-2 \cos(2 \pi n_{0}) (J_0(2 \pi n_x) - 1)
    +\cos^2 (2 \pi n_{0}) (J_0^2(2 \pi n_x)  -1)\Big].
\end{equation}
The results in Eqs.~(\ref{eq:acstark1}),~(\ref{eq:acstark12}) arise from a simplified model of treating the transmon as a charge qubit, although the parameter regime $E_J/ E_C$  of these two are in fact the opposite limits. However, the results are qualitatively correct, as seen in Fig.~\ref{fig:sstark025}. This is mostly due to the fact that the charge dispersion for the higher transition (see section \ref{sec:sdisp}) is relatively correctly predicted.

In the experiment, we also observe the effect of a quasiparticle which intermittently tunnels in and out of the island. The presence of the quasiparticle shifts the charging energy by half a period, see Fig.~\ref{fig:ChargingEnergy}a. Therefore, in a time-averaged measurement, two gate charge values differing by one electron appear as superimposed on top of each other. This behavior is rather typically observed in circuits where the single-electron charging effects are noticeable \cite{Aumentado}. As it is often the case in superconducting devices, the origin of the quasiparticles is somewhat unclear, but can be associated with radiation from higher-temperature parts of the experimental setup.

\begin{figure}
 \includegraphics[width=0.86\textwidth]{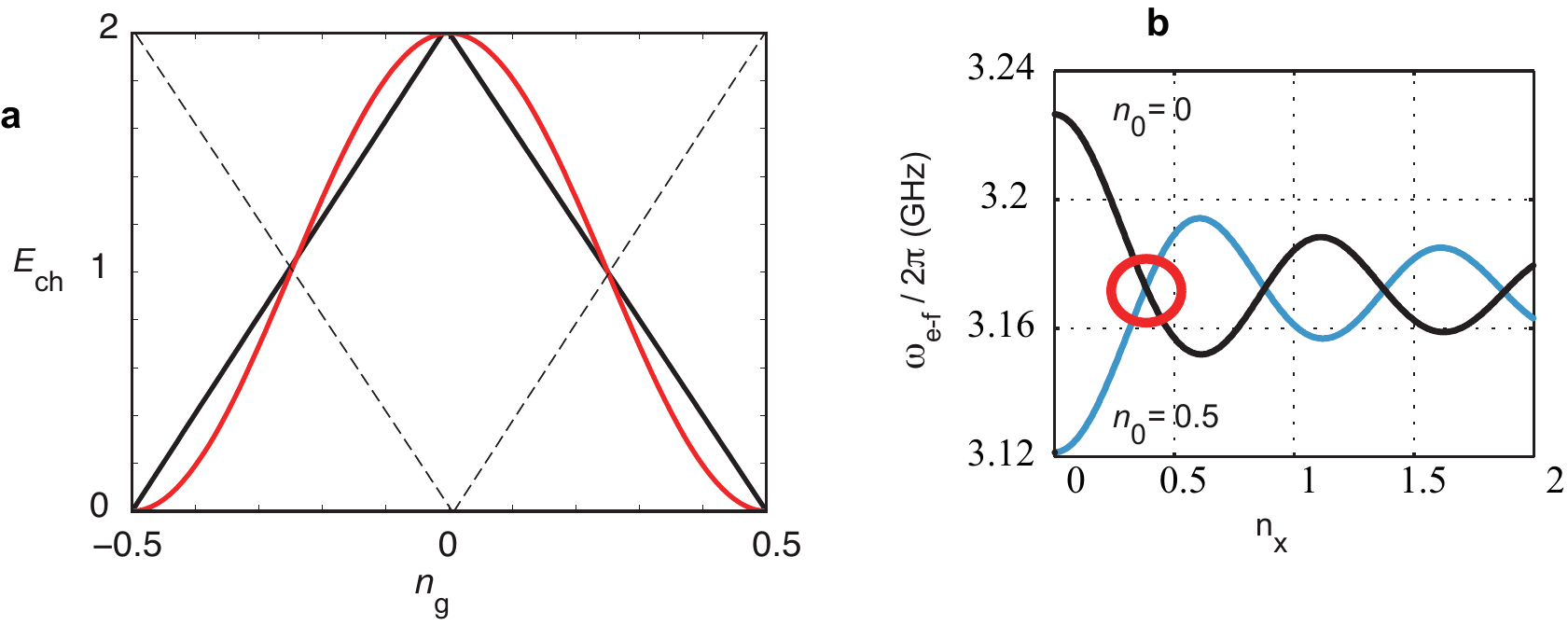}
  \caption{\textbf{Charging energy of the qubit}. \textbf{a}, Charging energy for the two lowest charge states in Eq.~(\ref{eq:H01}) (black solid line), and a sinusoidal approximation of it (red, Eq.~(\ref{eq:H01bsin})). The dashed lines depict the presence of an extra quasiparticle on the island. The y-axis is scaled with $E_C$ and x-axis is the gate charge. \textbf{b}, Illustration of a "dynamical sweet spot", where the qubit level spacing at a certain modulation amplitude is insensitive to changes in the offset charge $n_0$. The lines depicting the level spacings cross, independently of $n_0$, for the first time at about $ n_x \simeq 0.4$.}
  \label{fig:ChargingEnergy}
\end{figure}


\subsection{Using the charge dispersion of the transmon qubit}
\label{sec:sdisp}

The charge dispersion of the transmon is derived in the limit $E_J \gg E_C$ \cite{transmon}. Here, for convenience we use numeric labels $n=0,1,2,...$ for the eigenstates (not to be confused with number states used in sections \ref{shamilt}, \ref{sstark}). The eigenenergies are
\begin{equation}\label{eq:wntransmon}
 \omega_n(n_g) = \omega_n(n_g = 1/4) - \frac{\epsilon_n}{2} \cos(2\pi n_g) \,.
\end{equation}
The peak-to-peak amplitude is
\begin{equation}\label{eq:eps}
\epsilon_n = (-1)^n E_C \frac{2^{4n+5}}{n!}\sqrt{\frac{2}{\pi}} \LL(\frac{E_J}{ 2E_C}\RR)^{\frac{n}{2}+\frac{3}{4}} \exp\LL(-\sqrt{8E_J/E_C}\RR) \, .
\end{equation}
Since the higher levels have progressively stronger charge dispersion, a good approximation for the transition energies between two levels is given by $\omega_{\m{n-1,n}} \simeq  \omega_{n}$.

Let us take the gate dependence as a slow modulation of the level energy in the eigenbasis, and, for simplicity, label $\omega_0 \equiv \omega_n(n_g = 1/4)$. We use the previous notation for the Pauli matrices which operate now in the eigenbasis,
\begin{equation}
H_q = - \frac{1}{2} \LL[ \omega_0 + \epsilon_2 \cos \LL(2\pi n_g (t) \RR) \RR] \PauliZ  \label{Eq:ModelHeps}
\end{equation}
where $n_g(t)$ is given by Eq.~(\ref{eq:nx}). In the same way as starting from Eq.~(\ref{eq:H01bsin}) in the charge-qubit model in section \ref{sstark}, we use a Bessel expansion to break this into multiple frequencies. However, in contrast to the simplistic model, we now include all frequencies:
\begin{equation}
H_q = - \frac{1}{2} \delta_0 \PauliZ  - \frac{1}{2}  \sum_{\beta=1} \delta_{\beta} \cos \LL( \beta \omega_g t \RR) \PauliZ  \label{Eq:ModelHepse}
\end{equation}
where
\begin{equation}\label{eq:deltak}
\begin{split}
\delta_0 = & \omega_0  + \frac{\epsilon_2}{2}  \cos(2 \pi n_0)  J_{0}(2 \pi n_x)\\
\delta_{\beta} = &(-1)^{\beta/2} \epsilon_2   \cos(2 \pi n_0)  J_{\beta}(2 \pi n_x) , \m{\: \:\: \:  \: even} \: \beta \\
\delta_{\beta} = & (-1)^{(\beta+1)/2} \epsilon_2   \sin(2 \pi n_0)  J_{\beta}(2 \pi n_x) , \m{\: \: \:  odd} \: \beta
\end{split}
\end{equation}

\subsection{Spectroscopy by the excitation microwave}
\label{sstwotone}

\subsubsection{First sidebands}
\label{sstwotone1}

Let us consider as an introductory example a two-level Hamiltonian in its eigenbasis, coupled to \emph{two} linearly polarized fields:
\begin{equation}
H_q = - \frac{1}{2} \left[ \delta_0 + \delta \cos (\omega_g t) \right] \PauliZ - g_{\m{ext}} \cos (\omega_{\m{ext}} t) \PauliX . \label{Eq:ModelHamiltonian}
\end{equation}
The (low-frequency) field with frequency $\omega_g$ and Rabi frequency $\delta$ will describe the (strong) drive due to the phonon field, and the perpendicular field having the frequency $\omega_{\m{ext}}$ and Rabi frequency $g_{\m{ext}}$ models the excitation tone in two-tone spectroscopy. The probe tone used to measure the qubit state via dispersive cavity readout is not considered; it simply gives a phase shift proportional to the population of the upper level.

We move to a non-uniformly rotating frame with respect to the mechanical modulation by the transformation
\begin{equation}
u = \exp \left[ i \frac{\delta}{2 \omega_g} \sin (\omega_g t) \PauliZ \right].
 \label{Eq:ModelTransformation}
\end{equation}
We obtain
\begin{equation}
\begin{split}
H_q = & - \frac{ \delta_0}{2} \PauliZ - \sum_{k = -\infty}^{\infty} \Omega_{2k} \cos \LL[(\omega_{\m{ext}} + 2 k \omega_g)t\RR] \PauliX + \sum_{k = -\infty}^{\infty}  \Omega_{2k-1} \sin \LL[\omega_{\m{ext}} + (2k-1)\omega_g t\RR] \PauliY \,.
\label{Eq:ModelSB}
\end{split}
\end{equation}
The Rabi frequencies are
\begin{equation}
\begin{split}
 \Omega_k = g_{\m{ext}}J_k  \left( \frac{\delta}{\omega_g} \right) .
\label{Eq:DeltaRabi}
\end{split}
\end{equation}

\subsubsection{Strong drive}
\label{sstwotonen}

In reality the situation is more complex, since our two-level system (TLS) is coupled to a sum of fields having different frequencies, which are due to expanding the sine of sine (Eq.~(\ref{Eq:ModelHepse})), and also to the excitation tone:
\begin{equation}
H_q = - \frac{1}{2} \left[ \delta_0 + \sum_{\beta = 1}^{\infty} 
 \delta_{\beta} \cos (\beta \omega_g t) \right] \PauliZ - g_{\m{ext}} \cos (\omega_{\m{ext}} t) \PauliX .\label{Eq:Hsum}
\end{equation}
Here, one has to repeat the transformation $u$ of Eq.~(\ref{Eq:ModelTransformation}) separately for each frequency component, yielding the overall transformation:
\begin{equation}\label{UnonunibU}
\begin{split}
    U = \exp \LL[\frac{i \sigma_z}{2} \sum_{\beta=1}\frac{\delta_{\beta}}{\beta \omega_g} \sin(\beta \omega_g t)  \RR]  .
    \end{split}
\end{equation}
Equation (\ref{Eq:Hsum}) becomes
\begin{equation}
\begin{split}
H_q = - \frac{1}{2} \delta_0 \sigma_z - \frac{ g_{\m{ext}}}{2} \cos (\omega_{\m{ext}} t) (P + P^*) \sigma_x - \frac{i g_{\m{ext}}}{2} \cos (\omega_{\m{ext}} t) (P - P^*)\sigma_y , \\
\label{Eq:probeUU0}
    \end{split}
\end{equation}
where we used the shorthand notation
\begin{equation}
\begin{split}
 P \equiv \exp \LL[i\sum_{\beta=1}\frac{\delta_{\beta}}{\beta \omega_g} \sin(\beta \omega_g t) \RR] \, .
\label{Eq:probeUU}
    \end{split}
\end{equation}
Let us also define $ \delta'_{\beta} \equiv \frac{\delta_{\beta}}{\beta \omega_g} $. With further algebra we get
\begin{equation}
\begin{split}
P^* \Longrightarrow  & \sum_{p=-\infty}^{\infty} J_p \LL( \frac{\delta_1}{\omega_g}\RR) \exp( -ip\omega_g t) \times
\sum_{q=-\infty}^{\infty} J_q \LL( \frac{\delta_2}{2 \omega_g}\RR) \exp( -2iq\omega_g t) \times...\\
 = & \Big\{ ... + J_{-1} (\delta'_1) \exp( i \omega_g t) + J_0 (\delta'_1)  + J_1 (\delta'_1) \exp( -i \omega_g t) +J_2 (\delta'_1) \exp( -2i\omega_g t)  + ...\Big\} \times \\ 
  \times &
 \Big\{ ... + J_{-1} (\delta'_2) \exp(2 i \omega_g t) + J_0 (\delta'_2)  + J_1 (\delta'_2) \exp( -2 i \omega_g t) +J_2 (\delta'_2) \exp( -4i\omega_g t)  + ...\Big\} \times \\
 \times & ... 
    \end{split}
    \label{Eq:probeUU2} 
\end{equation}
We neglect sidebands of sidebands, since they have a lower amplitude. Then the explicitly written two lines in Eq.~(\ref{Eq:probeUU2}) yield multiples of the mechanical frequency up to third order:
\begin{equation}
 P^* \simeq J_0 (\delta'_1)J_0 (\delta'_2) - 2i J_1 (\delta'_1) J_0 (\delta'_2) \sin(\omega_g t) - 2i J_1 (\delta'_2) J_0 (\delta'_1) \sin(2 \omega_g t) - 2i J_1 (\delta'_3) J_0 (\delta'_2) \sin(3 \omega_g t) + ...
 \label{Eq:UUxm3}
\end{equation}
This result is applied in Eq.~(\ref{Eq:probeUU0}), obtaining
\begin{equation}
\begin{split}
H_q = - & \frac{1}{2} \delta_0 \sigma_z - \Omega_0  \cos (\omega_{\m{ext}} t)\sigma_x
-  \Omega_1 \LL[  \sin \LL( (\omega_{\m{ext}} + \omega_g ) t\RR)
-  \sin \LL( ( \omega_{\m{ext}} - \omega_g )t\RR)\RR]\sigma_y + \\
+& \Omega_2 \LL[ \sin \LL((\omega_{\m{ext}} + 2\omega_g )t\RR)
 - \sin \LL((\omega_{\m{ext}} - 2\omega_g )t\RR)\RR]\sigma_y
 + ...
\label{Eq:probeUUPm}
    \end{split}
\end{equation}
The Rabi frequencies are now
\begin{equation}\label{eq:DRabi}
\begin{split}
    \Omega_0 = & g_{\m{ext}} J_0 \LL( \frac{\delta_1}{\omega_g} \RR) J_0 \LL( \frac{\delta_2}{2 \omega_g} \RR)  \\
     \Omega_{ 1} = &  g_{\m{ext}} J_1 \LL( \frac{\delta_1}{\omega_g} \RR) J_0 \LL( \frac{\delta_2}{2 \omega_g} \RR)  \\
         \Omega_{ 2} = & g_{\m{ext}} J_1 \LL( \frac{\delta_2}{2\omega_g} \RR) J_0 \LL( \frac{\delta_1}{ \omega_g} \RR) \\
         \Omega_{ 3}= &  g_{\m{ext}} J_1 \LL( \frac{\delta_3}{3\omega_g} \RR) J_0 \LL( \frac{\delta_2}{2 \omega_g} \RR)  \,. \\
         \end{split}
\end{equation}
The higher Rabi frequencies ($k > 1$) are suppressed from those in Eq.~(\ref{Eq:DeltaRabi}), nearly comparable to that of the main peak ($k =0$), due to the nested Bessel function dependence.

\subsection{Level populations}

One can estimate the populations of the qubit levels in the spectroscopy experiment discussed in section \ref{sstwotone}. Let us consider a generic TLS described by a spin Hamiltonian, with a notation reminiscent of this discussion:
\begin{equation}\label{eq:ham}
    H_q = -\frac{\delta_0}{2} \sigma_z -\frac{\Omega_x}{2} \sigma_x -\frac{\Omega_y}{2} \sigma_y \,.
\end{equation}
This Hamiltonian is time-independent, however, in a rotating frame, a driven system will adopt similar form. For notational simplicity, let 0 and 1 denote the ground and first excited states of the TLS.


The Liouvillean master equation for the density matrix is:
\begin{equation}\label{eq:qbmeq}
    \dot{\rho} = \frac{1}{i \hbar }\LL[H,\, \rho \RR] + \mathcal{L}[\rho] \,,
\end{equation}
where the Liouvillean with decay and pure dephasing is:
\begin{equation}\label{eq:liouv}
    \mathcal{L}[\rho] = - \frac{\gamma}{2} \LL( \sigma^+ \sigma^- \rho
    + \rho \sigma^+ \sigma^- - 2\sigma^-  \rho \sigma^+ \RR)
    + \frac{\gamma_{\phi}}{2} \LL( \sigma_z \rho \sigma_z  -  \rho \RR) \,.
\end{equation}
%

%
%

The steady-state population of the excited state in the Rotating-Wave Approximation becomes
\begin{equation}\label{eq:rhooSS}
    \rho_{11} = 1-\rho_{00} = \frac{\frac{\gamma_{\m{tot}}}{2 \gamma}\LL( \Omega_x^2 +\Omega_y^2 \RR) }{ \delta_0^2 +\frac{\gamma_{\m{tot}}}{\gamma}\LL( \Omega_x^2 + \Omega_y^2\RR) +  \gamma_{\m{tot}}^2 }\,,
    \end{equation}
Here, the total decoherence rate is
\begin{equation}\label{eq:gammatot}
     \gamma_{\m{tot}} = \frac{\gamma}{2} + \gamma_{\phi} \, .
    \end{equation}
Equation~(\ref{eq:rhooSS}) is Lorentzian having the full width at half maximum
\begin{equation}\label{eq:rhooSSgen}
   \delta \omega = 2 \sqrt{\LL( \Omega_x^2 +\Omega_y^2 \RR) \frac{\gamma_{\m{tot}}}{\gamma}+  \gamma_{\m{tot}}^2} \,.
    \end{equation}
%
%
%

%
%


In order to apply these results for the Hamiltonians Eq.~(\ref{Eq:ModelSB}),~(\ref{Eq:probeUUPm}) describing the spectroscopy experiment, we make another transformation $u_2 = \exp \LL[-i \LL( \omega_{\m{ext}} + k \omega_g\RR) t \sigma_z /2  \RR]$ into a frame rotating with a particular sideband $k = -\infty ... \infty$. We can then use Eq.~(\ref{eq:rhooSS}) for finding the steady-state population. One next makes the $u_2$ transformation into the frame of each of the sidebands separately. If the peaks do not overlap, i.e., $\delta \omega \ll \omega_m$, one can sum the occupancies due to each such operation:
\begin{equation}
 \rho_{11} = \frac{1}{2}\sum_{k=-\infty}^{\infty} \frac{\gamma_{\mathrm{tot}} \Omega_k^2}{\gamma ( \delta_0 - \omega_{\m{ext}}  - k \omega_g)^2 + \gamma_{\mathrm{tot}} \left( \Omega_k^2 + \gamma  \gamma_{\mathrm{tot}} \right)} \,,
\label{eq:rhoo22}
\end{equation}
where the Rabi frequencies are given either by Eq.~(\ref{Eq:DeltaRabi}) for the case of weak drive, or, more generally by Eq.~(\ref{eq:DRabi}). In the latter case, the sideband Rabi frequencies are suppressed from those in Eq.~(\ref{Eq:DeltaRabi}) due to the nested Bessel function dependence.

Using these results, we can plot a prediction for the oscillatory Stark shift whose measurement data was shown Fig.~2d in the main text. We obtain an excellent agreement, as displayed in Fig.~\ref{fig:sstark025}. The mirrored image is missing in the simulation since we have not included the offset quasiparticle in the simulation. The sideband visibility, proportional to its Rabi frequency, undulates with increasing $n_x$ according to the nested Bessel function dependence.

\begin{figure}[h]
 \includegraphics[width=0.57\linewidth]{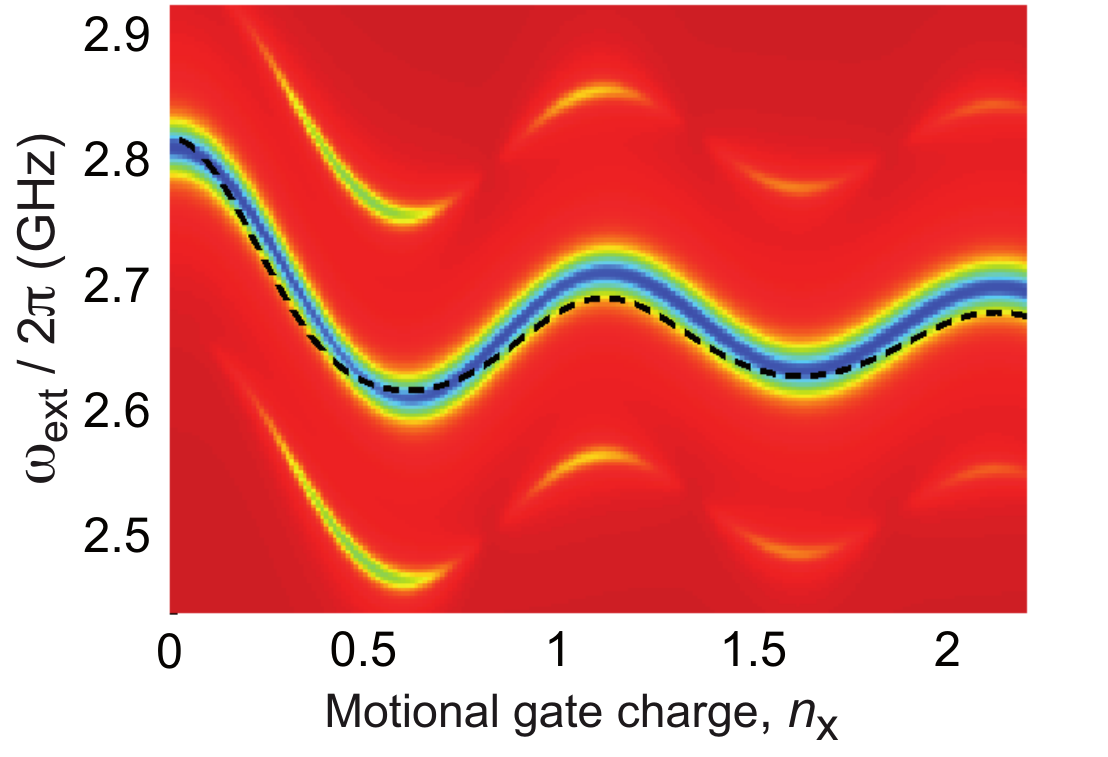}
  \caption{\textbf{Analytical model of Fig.~2d in main text}. The dashed line is the charge-qubit limit prediction, Eq.~(\ref{eq:acstark12}), however, with unrealistically low $E_J(\Phi_{\m{dc}}/\Phi_0 = 0.370) /2\pi = 6.0$ GHz. This is due to that the charge qubit model exaggerates the charge dispersion in the current transmon limit $E_J \gtrsim E_C$.}
  \label{fig:sstark025}
\end{figure}

\subsection{Floquet calculation of the driven transmon}
\label{sfloq}

According to the Floquet theorem \cite{ShirleyPR1965,SonPRA2009}, the time-dependent Schr{\"o}dinger equation,
\begin{equation}
H(t) | \Psi (t) \rangle = i \hbar \partial_t | \Psi (t) \rangle, \label{Eq:TimeDepSchr}
\end{equation}
with a time-periodic Hamiltonian, $H(t) = H(t+T)$, has a solution
\begin{equation}
| \Psi_{\alpha} (t) \rangle = | u_{\alpha} (t) \rangle e^{-i \varepsilon_{\alpha} t / \hbar}, \label{Eq:FloquetSolution}
\end{equation}
where $| u_{\alpha} (t) \rangle$ is periodic in time with period $T = 2 \pi / \omega$. Substituting Eq.~(\ref{Eq:FloquetSolution}) into Eq.~(\ref{Eq:TimeDepSchr}), we obtain the equation for the quasienergy state $| u_{\alpha} (t) \rangle$,
\begin{equation}
\left[H(t) - \varepsilon_{\alpha}  \right] | u_{\alpha} (t) \rangle = i \hbar \partial_t | u_{\alpha} (t) \rangle, \label{Eq:TransformedSchr}
\end{equation}
which is just the time-dependent Schr{\"o}dinger equation with the energy shifted by the quasienergy $\varepsilon_{\alpha}$.

We expand the Hamiltonian and the quasienergy state with their Fourier components as
\begin{align}
H(t) = & \sum_n \sum_{\alpha, \beta} h_{\alpha \beta}^n e^{in \omega t} | \alpha \rangle \langle \beta |, \\
| u (t) \rangle = & \sum_m \sum_{\alpha} c_{\alpha}^m e^{i m \omega t} | \alpha \rangle.
\end{align}
Plugging these into Eq.~(\ref{Eq:TransformedSchr}) yields
\begin{equation}
\sum_{m} \sum_{\beta} \left(  h_{\alpha \beta}^{n-m}  + \delta_{n,m} \delta_{\alpha,\beta} m \hbar \omega \right) c_{\beta}^m = \varepsilon c_{\alpha}^{n}.
\end{equation}
Defining the Floquet matrix $H_F$ as
\begin{equation}
\langle \alpha n | H_F | \beta m \rangle = h_{\alpha \beta}^{n-m} + m \hbar \omega \delta_{n,m} \delta_{\alpha,\beta},
\end{equation}
this eigenvalue problem can be written in a simple matrix form
\begin{equation}
H_F \vec{c} = \varepsilon \vec{c}.
\end{equation}
Thus, we have transformed the finite-dimensional time-dependent problem to an infinite-dimensional time-independent problem. However, the number of Floquet blocks can be truncated depending on the strength of the periodic drive leading to computable eigenvalue problem.

%
%

\begin{figure}[ht!]
 \includegraphics[width=0.85\linewidth]{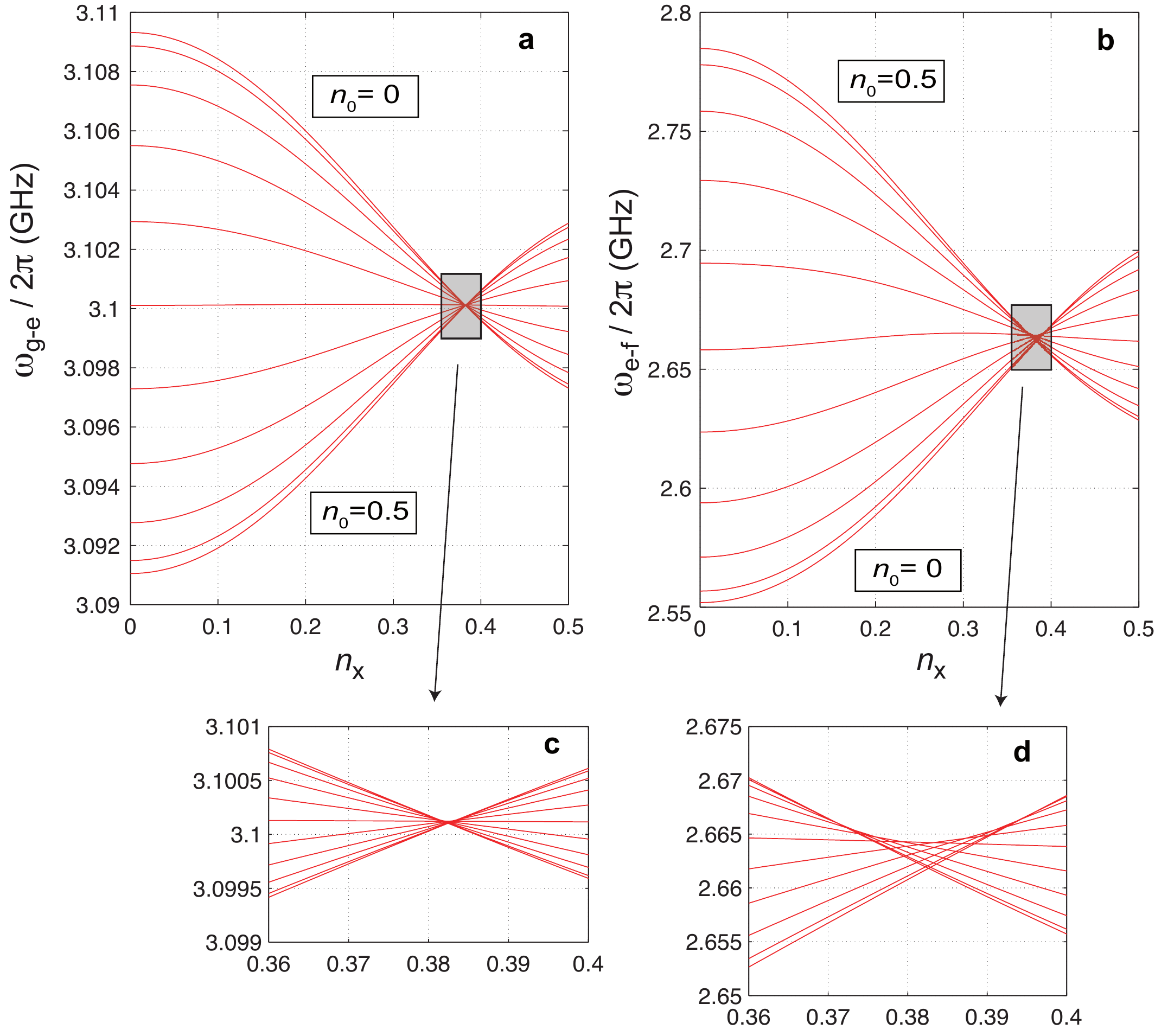}
  \caption{\textbf{Floquet results of the dynamical sweet spot}. \textbf{a}, Transition frequency between the ground state and first excited state, as a function of gate charge modulation amplitude. \textbf{b}, Transition between the first and second excited states. \textbf{c},\textbf{d}, Zoom-in of the regions marked in \textbf{a},\textbf{b}, respectively, where the transition is insensitive to charge offset. The family of curves depict changing $n_0$ in steps of 0.05 between 0 and $1/2$ as labelled. Notice the different scales in all panels.}
  \label{fig:ssweet}
\end{figure}

The Hamiltonian we consider is given in the Cooper pair basis by the qubit Hamiltonian (Eq.~(\ref{eq:cptHamiltn3})) with the gate charge driven sinusoidally, viz.~Eq.~(\ref{eq:nx}). This procedure allows for an exact treatment of the transmon qubit. The Fourier components of the Hamiltonian are
\begin{align}
H^{[0]} = & H_q, \\
H^{[\pm 1]} = & 4 E_C \left( n_0 - n \right) n_x.
\end{align}
The Floquet Hamiltonian is thus given by
\begin{align}
H_F = & \left( \begin{array}{ccccc}
\ddots & \ddots  &   &   &   \\
 \ddots     & H^{[0]} - \omega \hat{I}   & H^{[-1]}  &   &   \\
      & H^{[+1]}  & H^{[0]}  &  H^{[-1]} &   \\
      &   &  H^{[+1]} &  H^{[0]} + \omega \hat{I}  & \ddots  \\
      &   &   & \ddots  &  \ddots
\end{array} \right),
\end{align}
where the dimension of $H^{[0]}$ and $H^{[\pm 1]}$ is given by the number of Cooper pair number states used. In the modeling presented, 41 Cooper pair number states were used with 201 Floquet states. The size of the Floquet matrix was therefore $8241 \times 8241$.

The transition frequencies plotted in Fig.~2d in main text and in Fig.~\ref{fig:stark025}, are then obtained as a difference of a proper pair of quasienergies \cite{BSpaper}. In order to obtain a numerically exact solution for the qubit population, one could in principle include the qubit excitation tone in the Floquet analysis as well, however, the straightforward diagonalization becomes intractable.

\subsection{Dynamical sweet spot}
\label{sspot}

We can evaluate the dynamical sweet spot, mentioned in Fig.~\ref{fig:ssweet}, in some more detail. The qubit is coupled to a classical field Eq.~(\ref{eq:nx}) which can be, in particular, the mechanical resonator. The relevant part of the Hamiltonian for the problem of dephasing is that in Eq.~(\ref{Eq:ModelHeps}), repeated here:
\begin{equation}
H_{q}=
-\frac{\hbar}{2}\left[\omega_{0} + \frac{\epsilon_{2}}{2}\cos (2\pi n_{0})J_{0}(2\pi n_{x})\right] \sigma_{z}.
\label{eq:dynsweet1}
\end{equation}
Consider now a fluctuation of the parameters $n_{0},n_{x}$ around the mean values: $n_{0}=n_{0} +
\delta n_{0}$, $n_{x}=n_{x} + \delta n_{x}$. Using first order perturbation theory, we can get a simple expression for the pure dephasing rate $\gamma_{\phi}$ induced by these fluctuations. We assume that the noises in $n_{0}$ and $n_{x}$ are uncorrelated and we use $J_{0}'(z) = - J_{1}(z)$. We obtain $\gamma_{\phi}=
\gamma_{\phi, n_{0}} + \gamma_{\phi, n_{x}}$, where
\begin{eqnarray}
\gamma_{\phi, n_{0}} &=& \pi^3 \epsilon_{2}^2 \sin^2 (2 \pi n_{0}) J_{0}^{2}(2\pi n_{x}) S_{n_{0}}(0)  ,\\
\gamma_{\phi, n_{x}} &=& \pi^3 \epsilon_{2}^2 \cos^2 (2 \pi n_{0}) J_{1}^{2}(2\pi n_{x}) S_{n_{x}}(0) .
\end{eqnarray}
and $S_{n_{0}}(0)$, $S_{n_{x}}(0)$ are the corresponding zero-frequency noise powers.
This result shows clearly that at the dynamical sweet points (that is, the zeros of $J_{0}(2\pi n_{x}))$ the noise due to fluctuations of charge is suppressed. At the same time, it shows that the system is insensitive to fluctuations in the driving amplitude at certain amplitude spots (the zeroes of $J_{1}^{2}(2\pi n_{x}$)).

We also observe that at the dynamical sweet point, supposing Eq.~(\ref{eq:dynsweet1}) holds (we will evaluate this below) the system is immune to charge noise to any order. This contrast with the known case of the charge sweet point, which makes the system insensitive only to first-order fluctuations in the charge. There, second-order fluctuations are known to still produce a measurable effect on the dephasing time. It is worth pointing out that also the origin of these two effects is different.
In the case of the charge sweet point, the effect is local (when thinking in terms of the superconducting phase $\varphi$ as the coordinate), because it is due to the bending of the energy levels near the avoided crossover. This does not depend on the behaviour of the system at values of the phase further away from the crossover point. In particular, the phase periodicity $\varphi \rightarrow \varphi + 2\pi$ is irrelevant.
In the case of the dynamical sweet point, the $\cos$-bending of the energy levels is due to a nonlocal effect in the phase coordinate, namely the establishment of band-like structures due to tunneling between the adjacent wells of the Josephson potential $-E_{\rm J}\cos (\varphi)$. The periodicity $\varphi \rightarrow \varphi + 2\pi$ is essential, and the effect cannot be obtained by a perturbative expansion. Also in our case the modulation is realized with a large amplitude $n_x$, sweeping $n_{g}$ across a few wells.

Regardless of coupling to a mechanical resonator, the dynamical sweet spot can hence be a useful concept for suppressing a pure dephasing contribution to qubit decoherence in general. It can allow a transmon to operate more towards charge regime, which allows more anharmonicity. The principle could be the same way used to suppress flux noise, by applying a large-amplitude, slow flux drive. In order to suppress charge noise when coupling a charge-regime transmon to a mechanical resonator in the quantum limit, a slow "dummy" gate drive can be used to create these conditions.

The Floquet calculation provides an accurate account of the dynamical sweet spot. This takes into account the fact that towards the charge qubit limit, the energy levels deviate from sinusoidal shape. This leads to the fact that they no longer cross at exactly the same spot, as seen in Fig.~\ref{fig:ssweet} which is plotted for the present device. For the lowest transition, the set of curves corresponding to different offsets $n_0$ cross at the same point with 0.2 \% accuracy, whereas for the next higher transition, the remaining dispersion is about 1 \%. The enhancement of the dephasing time can be estimated based on the reduction of charge dispersion. For example, a 1 \% remainder would roughly correspond to a 100-fold increase of the dephasing time, and hence operations even in the charge qubit limit can be possible at arbitrary gate offsets.

\subsection{Electromechanical interaction in the transmon limit}
\label{sopt}

In this section, we will consider ways to increase the electromechanical coupling $g_m$. This will be beneficial in order to reach the quantum limit. The coupling, Eq.~(\ref{eq:transcavHamsg}), can be enhanced by (a) increasing $V_{\m{dc}}$, (b) increasing $E_C$, (c) increasing $dC_g/dx$. Let us fix $V_{\m{dc}} = 5$ Volts as presently. The simplest way is then to decrease the qubit capacitance $C_q$, that is, turning the transmon towards a charge qubit. Although this will make the qubit more susceptible for background charge noise, we can take advantage of the dynamical sweet spot (section \ref{sspot}) to suppress the noise.

Apart from the dynamical sweet spot (Figs.~\ref{fig:ChargingEnergy},~\ref{fig:ssweet}), the effect of background charge noise can be eliminated by operating the qubit in the full transmon regime with little charge dispersion, however, this requires a clearly larger $g_m$ than presently. As seen in Fig.~\ref{fig:qHamilt}, while the parallel (charge) coupling $g_{m,z}$ vanishes in the transmon limit, the transverse coupling $g_{m,x}$ becomes roughly equal to $g_m$. The transverse coupling is understood as a modulation of the qubit dynamical variables ($\varphi$, $\dot{\varphi}$) by the motion. This contrasts the parallel (radiation-pressure type) coupling, which directly affects the qubit energy. A small transverse coupling, however, in the far-dispersive limit has little effect. Let us consider a device roughly similar to the present, but with the resonator having a 20 nm vacuum gap. Here we consider the lowest $g-e$ transition. We set $C_q = 80$ fF, $E_J/2\pi = 10$ GHz, $E_J/E_C \simeq 45$ and thus vanishing charge dispersion. We also obtain $g_{m,x} /2\pi\simeq 25$ MHz, $g_{m,z} \ll g_{m,x}$. We discuss prospects of this design in section \ref{stransfer}.
We believe the 20 nm vacuum gap can be made using the same technique as presently. Or, we can use partial filling of the gap with dielectric material, a technique which has demonstrated 32 nm gaps \cite{Nguyen2008}.
 
Regardless the type of coupling (parallel or transverse), the computational operations on the qubit, and its coupling to mechanical resonator can be separated with a coupling switchable on and off over ns-timescales. This is accomplished by the aid of either one or both of the following: First, the coupling has the strongest effect at low qubit frequency, due to detuning-dependence. For example, one can run the qubit at say 7 GHz, and use typical ns-timescale flux "shift" pulses which tune the qubit frequency down to 2-4 GHz. Second, we notice that unless the qubit is driven on the sideband resonance, it is effectively decoupled from the mechanical resonator. Then, both qubit frequency and the drive can be tuned when turning the coupling on or off.

\vspace{1cm}
\subsection{Motional signatures in the quantum limit}
\label{slown}

While the present work focuses mostly on high phonon numbers, it is appealing to foresee the physics when $(N_m, N_m^T) \rightarrow 0$. We remind that by $N_m$ we mean the driven phonon occupancy, and $N_m^T$ is the equilibrium population due to finite temperature.


A possibility to observe zero-point motion is to study its Stark effect. Due to high detuning of qubit and phonons, the shift per phonon is small. For the $|e \rangle$-$|f \rangle$ transition, we expect shifts of the order 1 kHz ... 100 kHz per phonon in the quantum limit, see Fig.~\ref{fig:qstark}. While this is substantially smaller than the qubit peak width, one can inspect it by ÒsittingÓ at the slope of spectroscopy peak. The precision can be estimated from Fig.~3c in main text. Here, we obtained about 0.1 degrees precision at 1 seconds integration time per point, and maximum slope 15 degrees per 50 MHz. These translate into quite manageable integration times, proportional the second power of required precision, less than a few minutes in the present setup, to see a single phonon at flux biases between 0.4 ... 0.6.

Another option is to observe motional sidebands in this limit, similarly to trapped ions. To plot the qubit population, we use Eq.~(\ref{eq:rhoo22}) with the motional gate charge due to zero-point motion ($x \rightarrow x_{\m{zp}}$ in Eq.~(\ref{eq:nxdef})). We are not attempting here to make a full quantum analysis, and hence these results will change somewhat in the very quantum limit. Especially, the red $k=-1$ sideband is expected to vanish at zero temperature, while the present semiclassical analysis gives equal red and blue sidebands. Similarly, we neglect the associated cooling. However, we argue the semiclassical analysis provides otherwise an accurate account of the magnitude and observability of the motional sidebands.

In Fig.~\ref{fig:sqside}a we show the predicted qubit population when $(N_m, N_m^T) \rightarrow 0$, for the present sample, but at very high excitation tone irradiation to the $|e \rangle$-$|f \rangle$ transition. The first motional sidebands (two arrows) are basically visible with the present signal-to-noise, better in the derivative. We thus obtain that the zero-point limit should be accessible, but for the present sample would need high Rabi frequencies ($g_{\m{ext}}/2\pi \gtrsim 150$ MHz) which are experimentally impractical. However, decrease of qubit damping by an order of magnitude would render the quantum sidebands visible. In the charge qubit case with the same resonator, In Fig.~\ref{fig:sqside}b, the motional sidebands are clear even in the linear regime of excitation tone.

A slight decrease of the qubit capacitance from the present value will render the low-phonon-number sidebands clearly visible.

\begin{figure}[ht!]
 \includegraphics[width=0.6\linewidth]{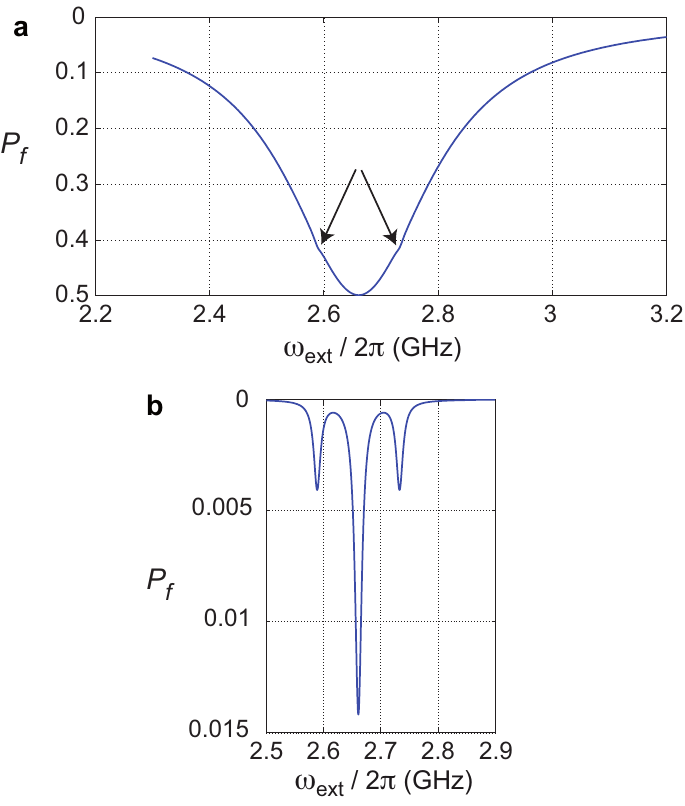}
  \caption{\textbf{Motional sidebands in the quantum limit (theory)}. Semiclassical two-level model, Eq.~(\ref{eq:rhoo22}), zero temperature limit. Relaxation to ground state is ignored. \textbf{a}, The qubit $|f \rangle$ state population for the present sample, but at a high (main peak) Rabi frequency $g_{\m{ext}}/2\pi = 138$ MHz. \textbf{b} same for a charge qubit, $E_J/E_C = 1$, $g_m/2\pi = 70$ MHz. Rabi frequency $g_{\m{ext}}/2\pi = 1.4$ MHz. Note that the equal appearance of the sidebands is due to the semiclassical approximation.}
  \label{fig:sqside}
\end{figure}

\subsection{State transfer using sideband operations}
\label{stransfer}

The measured sideband Rabi oscillations show the swapping of qubit excitations into phonons and back, in the classical regime. An interesting question is then, how well the qubit state can be swapped to the mechanical resonator in the quantum regime.

\begin{figure}[h]
 \includegraphics[width=0.95\linewidth]{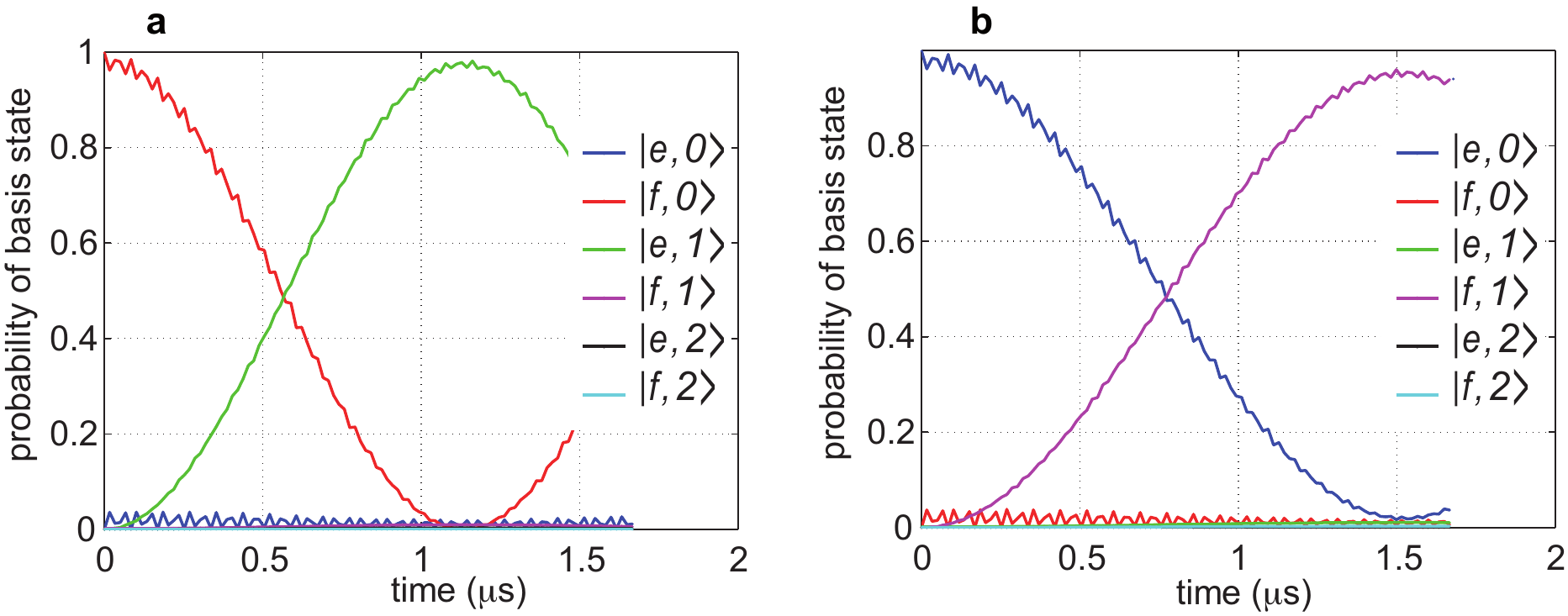}
  \caption{\textbf{Sideband state transfer in the quantum limit}. Time-domain simulation of Eq.~(\ref{eq:sHd}) at zero temperature, and decoherence neglected. The parameters correspond to the present experiment, $n_0 \simeq 0.25$,  $\Phi_{\m{dc}}  / \Phi_0 \simeq \pm 0.4$, $\Omega_x/2\pi = 10$ MHz, $\Omega_z/2\pi = 25$ MHz. \textbf{a}, red sideband, $-\omega_{\m{e-f}} + \omega_{\m{ext}} = -0.98\, \omega_m$. \textbf{b}, blue sideband, $-\omega_{\m{e-f}} + \omega_{\m{ext}} = 0.975 \,\omega_m$.}
  \label{fig:stransfer}
\end{figure}

We consider the model in Eq.~(\ref{eq:sH}), plus a general drive to the qubit,
\begin{align}
    H = -\frac{\omega_{\m{e-f}}}{2} \sigma_z  + \omega_m (b^{\dagger} b + 1/2)  + g_{m,z}  ( b^{\dagger} + b) \sigma_z + g_{m,x}   ( b^{\dagger} + b) \sigma_x 
    + \cos ({\omega_{\m{ext}} t})  \LL( \Omega_z \sigma_z + \Omega_x \sigma_x \RR)\,. 
    \label{eq:sHd}
\end{align}
%
%
%
In a uniformly rotating frame, and after rotating-wave approximation, we obtain
\begin{equation}
H = \frac{1}{2} \LL (-\omega_{\m{e-f}} + \omega_{\m{ext}} \RR) \sigma_z  +  \frac{\Omega_x}{2}  \sigma_x +\omega_m (b^{\dagger} b + 1/2)  + g_{m,z}  ( b^{\dagger} + b) \sigma_z 
\end{equation}
Hence, when $\sqrt{(-\omega_{\m{e-f}} + \omega_{\m{ext}})^2 + \Omega_x^2} = \pm \omega_m$, the driven qubit and the mechanical resonator become resonant and can efficiently exchange single quanta. The transfer time is given by the effective transverse coupling $g^* = \frac{\Omega_x}{\omega_m} g_{m,z}$ in the driven qubit eigenframe. In Fig.~\ref{fig:stransfer} we give examples of simulated sideband processes, for qubit prepared either in the ground state or excited state in the beginning. The transfer fidelity is limited by a corresponding residual diagonal coupling, however, we can obtain 97 \% fidelity.

Figure S14 displays the red-sideband state transfer for a transmon-regime qubit, as described in section \ref{stransfer}. This shows good prospect of coupling to the mechanical resonator also in the limit where background charge noise vanishes.

\begin{figure}[h]
 \includegraphics[width=0.45\linewidth]{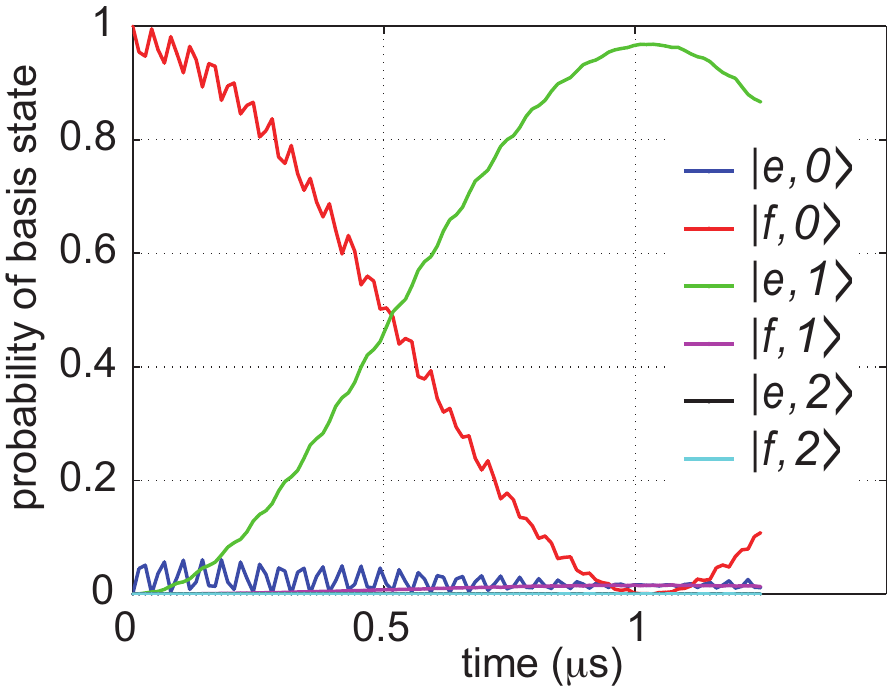}
  \caption{\textbf{Sideband state transfer in the quantum limit}. The parameters are selected for \emph{full transmon limit}, and we consider here the lowest transition $|g\rangle$-$|e\rangle$. The parameters are $g_m/2\pi = 25$ MHz, $\omega_{\m{g-e}}/2\pi = 4$ GHz, $\Omega_z/2\pi = 100$ MHz, $\Omega_x/2\pi = 20$ MHz, $-\omega_{\m{e-f}} + \omega_{\m{ext}} = -0.958\, \omega_m$, arbitrary $n_0$.}
  \label{fig:translim}
\end{figure}


\subsection{Sideband cooling}
\label{scool}

The motional sidebands correspond to the transitions which are frequently used to cool the motion of ions or mechanical resonators. For the latter, cooling has recently been actively experimented by coupling the motion to a linear cavity resonator. There has been theoretical work for using a two-level system instead of a linear cavity to cool mechanical systems, but little experimental effort. Two-level systems can be beneficial, since they offer orders of magnitude larger coupling, and hence stronger cooling at a given input microwave power. Ground-state cooling should be achievable starting from thermal occupancies $N_m^T \lesssim 10^2$ \cite{ZollerPRB2004,WallquistCool,RablPRB2010}.

For sideband cooling, the qubit excitation tone is applied at the first red sideband, $k=-1$. The absorption of a photon and emission of a phonon then together dissipate the energy inside the qubit. A distinction to regular cavity cooling is that the qubit has a finite spectrum, and thus saturation phenomena can take place. The cooling can be bottlenecked either by
the onset of coherent oscillations, or too slow dissipation in the qubit. In the latter case, if the resonator has occupation $N_m$ due to thermal bath or drive, cooling becomes strongly suppressed if quanta are introduced into the resonator at a rate faster than they can be dissipated, that is, if $N_m \omega_m/Q_m  > \gamma$. In the former case, detailed analysis tells \cite{RablPRB2010} that cooling becomes irrelevant when $N_m \gtrsim Q_m \frac{\Omega_0^2}{\gamma_{\phi} \omega_m }$. Overall, we obtain that  we can neglect cooling and heating if $N_m > 10^2 ... 10^3$, that is, in the present experiment.


\subsection{Effects of mechanical driving on qubit relaxation}

Since any additional component in the setup in principle opens a new relaxation channel for the qubit, one should consider losses within the mechanical resonator from this point of view.  A quick argument can be made for the case in which the number of mechanical quanta is very small. The direct relaxation rate via the mechanical resonator must be extremely small, because the qubit is highly detuned the oscillator, and its decay rate $\gamma_m/2\pi \approx 13$ kHz is very small, too. Indeed, the Purcell spontaneous emission rate from the excited level $f$ of the qubit to $e$ in the case of zero phonons is
$\gamma_P \simeq \gamma_m g_{m}^{2}/\Delta^2$, completely negligible. Here, $\Delta \simeq \omega_{\m{e-f}}$ is the detuning between the qubit and mechanical resonator, Eq.~(\ref{eq:tanThetaJC}). This argument is however not {\it a priori} valid in the case when the resonator is driven. In this case the qubit is dressed with a very large number of photons $N_{m}$ that can reach values of up to $\approx 10^6$.
As a result, the effective coupling increases by $g_{m}\rightarrow \sqrt{N_{m}}g_{m}$ and the loss rate of a phonon becomes as $ N_{m}\gamma_m$. This would seem to imply $\gamma_P \rightarrow N_{m}^2\gamma_P $, and would yield the Purcell decay rate $\gamma$ of the order of hundreds on MHz, making any
quantum-coherent operation impossible at large drivings. 

This is fortunately not the case, and a more careful calculation shows that the effect of driving is negligible even at high $N_{m}$'s. The proof assumes that the nanomechanical resonator remains harmonic even at large phonon numbers, and the resulting small effect is the result of coherent summation and destructive interference of decay paths in the first orders in $N_{m}$.
Consider an interaction Hamiltonian between the mechanical resonator and the bath,
\begin{equation}
H_\m{{m-env}} = \hbar \sum_{k}\lambda_{k}[b^{\dag}\chi_{k} + b\chi_{k}^{\dag}],
\end{equation}
where $\lambda_{k}$ is the coupling constant with the modes $\chi_{k}, \chi_{k}^{\dag}$ of the bath. 
We then apply the standard Fermi Golden rule arguments, by considering the decay of an initial state $i$ into a final state $f$ due to the loss of a single mechanical quanta into the environment with density of states  $\rho_{k}$
\begin{equation}
\gamma_{f\rightarrow i} = \frac{2\pi}{\hbar} \sum_{k}\rho_{k} \vert \langle 0, i|H_\m{{m-env}}|1, f\rangle\vert^{2}.
\end{equation}
If the (bare) nanomechanical decay rate is defined as $\gamma_m= 2\pi \hbar \sum_{k}\rho_{k}|\lambda_{k}|^2$, we obtain 
\begin{equation}
\gamma_{f\rightarrow i} = \gamma_m|\langle i|b^{\dagger}|f\rangle|^2 .
\end{equation}
We now take as the initial and final states the Jaynes-Cummings states. We consider transitions between the manifolds $N_{m}$ and $N_{m}-1$, corresponding to the emission of a quanta in the environment. For $N_{m} \gg 1$, by examining the structure  of these states, one sees that,  the transitions $|+, N_{m}\rangle \rightarrow |+, N_{m}-1\rangle$ and $|-, N_{m}\rangle \rightarrow |-, N_{m}-1\rangle$ do not modify significantly the occupation probability of the qubit. One can regard these transitions as due to the loss of phonons directly from the driving field, since they only reduce the phonon-number but they do not change significantly the weight of the
states $e$ and $f$. Thus the relevant transitions are the sideband transitions between the manifolds, $|+, N_{m}\rangle \rightarrow |-, N_{m}-1\rangle$ and $|- N_{m},\rangle \rightarrow |+, N_{m}-1\rangle$. To calculate these transition rates, we will use the approximation
\begin{equation}
1 + \frac{4g_{m}^2 (N_{m}+1)}{\Delta^2} \gg \frac{4g_{m}^2}{\Delta^2},
\end{equation}
which is clearly valid both in the case of small and large $N_{m}$ due to the fact that in our system $\frac{4g_{m}^2}{\Delta^2}\ll 1$.
We then calculate the quantities  $|\langle +, N_{m}|b^{\dagger}|-,N_{m}-1\rangle|^2$ and $|\langle -, N_{m}|b^{\dagger}|+,N_{m}-1\rangle|^2$. As already hinted before, these quantities will contain terms quadratic in $N_{m}$, which would result in $\gamma_P \simeq N_{m}^2 \gamma_m$. It turns out however that these terms cancel out, as well as the next-order terms (linear in $N_{m}$). We are then left with the following simple result for the Purcell effect, valid also for large values of $N_{m}$,
\begin{equation}
|\langle \pm, N_{m}|b^{\dagger}|\mp,N_{m}-1\rangle|^2 = \frac{1}{2}\frac{g_{m}^2}{\Delta^2} \frac{1}{1 + \frac{4g_{m}^2(N_{m}+1)}{\Delta^2}} \left[1 \pm \frac{1}{\sqrt{1 + \frac{4g_{m}^2 (N_{m}+1)}{\Delta^2}}}\right] \label{res}
\end{equation}
In the limit in which the phononic dressing is small, that is $\frac{g_{m}^2N_{m}}{\Delta^2} \ll 1$ we recover the known result
for the Purcell-effect change of the decay rate,
\begin{equation}
|\langle +, N_{m}|b^{\dagger}|-,N_{m}-1\rangle|^2 \approx \frac{g_{m}^2}{\Delta^2}.
\end{equation}
The result Eq. (\ref{res}) shows that relaxation rates associated with the motion of the mechanical oscillator are completely negligible, even when the dressing of the qubit is significant.

\vspace{0.5cm}

\subsection{Coupling to macroscopic quartz resonators}
\label{spiezo}

As mentioned in the end of main text, millimeter-sized quartz resonators have shown exceedingly high mechanical Q-values at cryogenic temperatures \cite{highQBAW0,highQBAW}. These are intriguing objects to connect to the present circuit QED setup.

We first consider our circuit to be patterned on top of the quartz chip itself. This creates a galvanic contact to the resonator, and the analysis is that of e.g., Ref.~\cite{ClelandPRA05}, except that we have small-capacitance junctions which allows for coupling to the background charge rather than junction voltage. We suppose electrodes on one side of the chip are formed by one of the qubit islands, and hence motion modulates the charge as in the present setup. Straightforward calculation gives the qubit-mechanical resonator coupling energy $g_m = 2 \frac{e_h}{\epsilon} e x_{\m{zp}}$. Here, $e_z$ is the shear piezoelectric modulus.

We take dimensions from Ref.~\cite{highQBAW}; the effective resonating volume 5 mm x 5 mm x 1 mm, $e_z = 0.1$ C/$m^2$, obtaining a coupling $g_m/2\pi \sim 0.3$ MHz which is lower than in the present experiment, but promising.

One can also maintain the spirit of contactless measurement by fabricating the circuit on a separate chip which is connected to the quartz resonator capacitively through a vacuum gap.

\subsection{Driven response of the mechanical resonator}
\label{sdriven}

With both dc and ac voltage applied to the mechanical resonator
\begin{equation}\
V_g (t) = V_{\m{dc}} + V_{\m{ac}} \cos(\omega_g t) ,
\end{equation}
 the driving force acting on it is
\begin{equation}\label{eq:fac}
    F = \frac{d}{dx} \LL( \frac{1}{2} C_g(x) V_g^2\RR) = |F|  \cos(\omega_g t) ,
\end{equation}
where
$$
 |F| = V_{\m{dc}} V_{\m{ac}}  \frac{d C_g}{dx} \, .
$$
The induced motion is
\begin{equation} \label{eq:SHOxpsol}
    x(t) =  \frac{ |F| \cos(\omega_g t + \Theta)}{ m \sqrt{ \omega_g^2 \omega_m^2 /Q_m^2 + \LL(\omega_g^2 - \omega_m^2\RR)^2}}
    \equiv |x| \cos(\omega_g t + \Theta) ,
\end{equation}
where
\begin{equation} \label{eq:SHOxpsolTheta}
    \tan \Theta = \frac{\omega_m \omega_g}{Q_m \LL( \omega_m^2 - \omega_g^2\RR) }  \, .
\end{equation}

The gate charge due to the applied $V_g$ is
\begin{equation} \label{eq:ngtotal}
\begin{split}
     n_g(t) = & \frac{C_g (t) V_g(t)}{2e} \simeq n_0 + \frac{d C_g }{d x} \frac{V_{\m{dc}} }{2e } |x| \cos(\omega_g t + \Theta) + \frac{C_g V_{\m{ac}}}{2e} \cos(\omega_g t)\\
   \equiv & n_0 +n_x\cos(\omega_g t + \Theta) + n_v \cos(\omega_g t)
   = n_0 + r \cos (\omega_g t + \theta)
    \, .
\end{split}
\end{equation}
which has been expressed as a single sinusoid on the last line with
\begin{align}
r 
= & \sqrt{ n_{v}^2 +  n_{x}^2 + 2  n_{v}  n_{x} \cos \Theta} \,,
\end{align}
and
\begin{equation}
\tan \theta = - \frac{ n_{x} \sin \Theta}{ n_{v} +  n_{x} \cos \Theta}.
\end{equation}

The phase shift $\Theta$ of motion with respect to the driving voltage appears as an asymmetry in the Stark shifts in Fig.~2 in the main text. 


One also obtains a relation between the motional gate charge and phonon number $N_m =  m \omega_m x_{\m{rms}}^2/\hbar$,
\begin{equation} \label{eq:snm}
  N_m 
  = \LL( \frac{4 E_C}{ \hbar g_m }\RR)^2 n_x^2 \, .
\end{equation}

The Stark shift, Eq.~(2) in the main text, is at low phonon number
\begin{equation} \label{eq:sstarknm}
       \Delta \omega_{\m{e-f}}/2 \pi \simeq \frac{\epsilon_{\m{e-f}}}{2} \pi^2 \cos(2 \pi n_0) \LL( \frac{ \hbar g_m }{4 E_C}\RR)^2 N_m \, .
\end{equation}
%


\end{widetext}

\end{document}